\documentclass[12pt,preprint]{aastex61}
\usepackage{amsmath,amssymb}
\usepackage{float}
\usepackage{color}

\def\deg{\ifmmode^\circ\else$^\circ$\fi}
\def\degc{\ifmmode^\circ\hbox{C}\else$^\circ$C\fi}
\def\arcsec{\ifmmode^{\prime\prime}\else$^{\prime\prime}$\fi}
\def\arcmin{\ifmmode^{\prime}\else$^{\prime}$\fi}

\def\earth{\ifmmode_{\mathord\oplus}\else$_{\mathord\oplus}$\fi}

\shorttitle{ATLAS Design and Hardware}
\shortauthors{Tonry et al.}

\begin{document}

\title{ATLAS: A High-Cadence All-Sky Survey System}

\author{J. L. Tonry}
\affiliation{Institute for Astronomy, University of Hawaii, 2680 Woodlawn Drive, Honolulu, HI 96822}
\email{tonry@hawaii.edu}
\author{L. Denneau}
\affiliation{Institute for Astronomy, University of Hawaii, 2680 Woodlawn Drive, Honolulu, HI 96822}
\email{denneau@hawaii.edu}
\author{A. N. Heinze}
\affiliation{Institute for Astronomy, University of Hawaii, 2680 Woodlawn Drive, Honolulu, HI 96822}
\email{aheinze@hawaii.edu}
\author{B. Stalder}
\affiliation{LSST, 950 N. Cherry Ave, Tucson, AZ 85719}
\email{bstalder@lsst.org}
\author{K. W. Smith}
\affiliation{Astrophysics Research Centre, School of Mathematics and Physics, Queen's University Belfast, Belfast, BT7 1NN, UK}
\email{k.w.smith@qub.ac.uk}
\author{S. J. Smartt}
\affiliation{Astrophysics Research Centre, School of Mathematics and Physics, Queen's University Belfast, Belfast, BT7 1NN, UK}
\email{s.smartt@qub.ac.uk}
\author{C. W. Stubbs}
\affiliation{Department of Physics, Harvard University, Cambridge, MA 02138, USA}
\email{stubbs@g.harvard.edu}
\author{H. J. Weiland}
\affiliation{Institute for Astronomy, University of Hawaii, 2680 Woodlawn Drive, Honolulu, HI 96822}
\email{hweiland@hawaii.edu}
\author{A. Rest}
\affiliation{Space Telescope Science Institute, 3700 San Martin Drive, Baltimore, MD 21218, USA}
\affiliation{Department of Physics and Astronomy, Johns Hopkins University, Baltimore, MD 21218, USA}
\email{arest@stsci.edu}

\begin{abstract}
Technology has advanced to the point that it is possible to image the
entire sky every night and process the data in real time.  The sky is
hardly static: many interesting phenomena occur, including variable
stationary objects such as stars or QSOs, transient stationary objects
such as supernovae or M dwarf flares, and moving objects such as
asteroids and the stars themselves.  Funded by NASA, we have designed
and built a sky survey system for the purpose of finding dangerous
near-Earth asteroids (NEAs).  This system, the ``Asteroid Terrestrial-impact Last Alert
System'' (ATLAS), has been optimized to produce the best survey
capability per unit cost, and therefore
is an efficient and competitive system for finding potentially 
hazardous asteroids (PHAs) but also for tracking variables and finding transients. 
While carrying out its NASA mission, ATLAS now discovers more bright ($m < 19$) 
supernovae candidates than any ground based survey, frequently detecting 
very young explosions due to its 2 day cadence. ATLAS discovered the afterglow
of a gamma-ray burst independent of the high energy trigger and has 
released a variable star catalogue of 5$\times10^{6}$ sources. 
This is the first of a
series of articles describing ATLAS, devoted to the design and
performance of the ATLAS system.  Subsequent articles will describe in more detail the software, the survey strategy, ATLAS-derived NEA population statistics, transient detections, and the first
data release of variable stars and transient lightcurves. 
\end{abstract}

\keywords{surveys; minor planets, asteroids: general; stars: variables: general; supernovae: general}

\section{Introduction}



The remarkable progress of silicon technology in recent decades has
made it possible to examine the entire sky for moving, variable, or
transient objects every night to a meaningful depth.  Optimizing
survey performance is a complex task, however. Resources need to be
divided between the cost of a facility to protect the system from the elements; 
a telescope, a mount, and a detector to collect the light; and computers, operations, and software to run the
survey and process the results.  Any of these features can limit
performance.

Combining the detector technology advances with venerable Schmidt telescopes
or newly designed wide-field facilities has rapidly changed the astronomical survey landscape in the last few years. 
The ambitious Pan-STARRS1 survey \citep[PS1;][]{2016arXiv161205560C}
has mapped 3$\pi$ steradians of the sky (30,000 square degrees) in 6 wavebands and is 
having a major impact --- not only in transients and moving objects 
\citep{2012ApJ...748L..15H,2014ApJ...795...44R}
but  from low mass stars \citep{2013ApJ...777L..20L}, through Milky Way stellar populations 
\citep{2015ApJ...813...44L} 
to the highest redshift quasars \citep{2014AJ....148...14B}.
The  Palomar Transient Factory \citep[PTF; ][]{2009PASP..121.1395L} 
reinvigorated the scientific capability of the Palomar Schmidt telescope producing 
a wide range of discoveries of novel objects 
\cite[e.g.][]{2011Natur.474..487Q,2011ApJ...736..159G,2013ApJ...769..130C}. 
PTF has been upgraded to the Zwicky Transient Facility (ZTF) \citep[][]{2014htu..conf...27B}
with a much larger field of view.
The QUEST camera was installed on the Schmidt telescope at La Silla 
to run the La Silla QUEST survey, \citep[LSQ;][]{2013PASP..125..683B} which 
combined with the the Public ESO Spectroscopic Survey
of Transient Objects  \citep[PESSTO][]{2015A&A...579A..40S} for spectroscopic follow-up, again producing a range of 
discoveries  \citep[e.g.][]{2014MNRAS.444.2096N,2015ApJ...807L..18N}.  
The  Catalina Real Time Survey \citep[CRTS;][]{2009ApJ...696..870D} is a very successful 
time-domain survey which has influenced survey science from the solar system through 
supernovae and AGN variability.  
The SkyMapper survey \citep{2007PASA...24....1K} is 
now producing its first public data products, completing the multi-color coverage of the 
whole sky \citep{2018arXiv180107834W}.  
Other surveys on large aperture telescopes such as the Dark Energy Survey 
\citep{2016MNRAS.460.1270D}
and HyperSuprimeCam 
\citep{2018arXiv180108240M}
are now playing a major role with exceptional depth and photometric 
performance over smaller sky areas. 
At the other end, novel use and 
fast processing of data from small 14cm lens systems by the  
All-Sky Automated Survey for SuperNovae \citep[ASSASN;][]{2017MNRAS.464.2672H}
have been impressively productive, providing some rare and surprising finds 
\citep{2016Sci...351..257D}. 
From the tens of centimeters to 10m sized apertures, 
survey astronomy truly has changed in the last few
years; a revolution that has made it into orbit with
ESA's Gaia facility using its scanning capability to produce transient
alerts \citep{2013RSPTA.37120239H,2015IAUGA..2257872W,2016A&A...595A...1G}.

ATLAS was proposed as a replicable system that NASA could use to find dangerous
asteroids, and optimization for the NASA mission opens synergistic
opportunities for many other types of science 
\citep{2011PASP..123...58T}.
Predicting
asteroid collisions with Earth places constraints on system capability,
for example, warning of at least one day for a $\sim$1~Mton explosion
requires all-sky monitoring at a sensitivity of $m>19$.  Funded in
2013, ATLAS achieved first light in June 2015 and now consists of two
independent units, one on Haleakala (HKO), and one on Mauna Loa (MLO) in 
the Hawai`ian islands. 

A number of papers have been written about sky survey design and optimization
including 
\cite{2011PASP..123...58T}, 
\cite{2011AN....332..714T,2016AJ....152..121T}, 
and
\cite{2016PASP..128h4501B}. 
\cite{2011PASP..123...58T}
summarized survey performance in terms of a ``survey speed''
that expresses the rate at which objects can be observed
to a limiting magnitude $m$ with signal to noise ratio SNR.  In the
background limited, random distribution on sky, Poisson regime this becomes

\begin{equation}
 SS = \frac{A\; \Omega_0 \; \epsilon \; \delta}{\omega}\; 10^{+0.4(\mu+m_0-2m_s)} = \frac{\hbox{SNR}^2 \; \Omega \;}{t_{cad}} \; 10^{+0.8(m-m_s)}
\label{eq:etendue}
\end{equation}

where $A$ [m$^2$] is the collecting area, $\Omega_0$ [deg$^2$] is the
solid angle covered by the detector, $\epsilon$ is the efficiency for
light to be detected (relative to a fiducial $m_0 = 25.10$ that
provides 1 photon per second per m$^2$ per 0.2 in natural log of
bandpass), $\delta$ is the duty cycle over cadence time $t_{cad}$ [sec]
that the shutter is open, $\omega$ [arcsec$^2$] is the point spread function (PSF) noise
footprint solid angle (essentially $3.5d^2$ where $d$ is the PSF
full width half maximum, FWHM), $\mu$ is the sky brightness [mag/arcsec$^2$], $\Omega$
[deg$^2$] is net solid angle surveyed during $t_{cad}$, and $m_s$ is a desired
survey depth.  In effect, 
the left hand side of equation~\ref{eq:etendue} describes a survey in design,
how well it ought to perform, the right hand side describes a survey
in operation, how well it actually performs.

Because this equation describes an extensible quantity, it is possible
to examine tradeoffs, such as doubling the collecting area or building
two identical systems in order to double the rate at which objects can
be found.  Less obvious trades that double the rate include halving
the PSF footprint solid angle $\omega$, looking for objects that are
0.4~mag brighter, or lowering the SNR requirement.
Since the bottom line for many surveys is how many objects can be
surveyed per unit time and the bottom line for any project is how
productive it is per unit resource, the metric by which a survey
project should be judged is survey speed per unit cost.

\cite{2011PASP..123...58T}
 showed that an array of 0.25~m astrographs could
inexpensively observe all sky each night to $m\sim19$.
The actual implementation of the
funded ATLAS program employs 0.5~m Schmidt telescopes rather
than an array of smaller telescopes, in order to
optimize with respect to telescope and detector cost --- advent of 10k CCDs brought the marginal cost of telescope improvement below that of detector improvement relative to the original design.

The relevant design numbers for one ATLAS unit are 
$A=0.14$~m$^2$ (including vignetting), 
$\Omega_0=29$~deg$^2$,
$\epsilon=1.25$ for transmission through atmosphere, optics,
detector QE, and $o$ filter bandpass width,
$\delta=0.75$, 
$\omega=52$~arcsec$^2$ for 2 pixel FWHM PSF,
$\mu=20.7$~mag/arcsec$^2$, and $m_s=20$, for a predicted speed of $SS=15$~deg$^2$/sec.
What an ATLAS unit can actually achieve in a 30~sec
exposure and 40~sec cadence in $c$ or $o$ band is $m_{5\sigma}\sim19.7$, 
corresponding to a ``best condition'' survey speed of
$SS=10$~deg$^2$/sec, less than Eq.~\ref{eq:etendue} because Eq.~\ref{eq:etendue} does
not include the terms for read noise or dark current.  While survey speed
might be made more useful by using median limiting magnitude and including a term for
weather losses (diminishing the ATLAS $SS$ to 3~deg$^2$/sec),
evaluating whether a system is performing to its design potential requires best
case numbers.  An ATLAS unit costs \$1M to replicate
(including everything from enclosure to software), so this ratio of survey speed to cost sets
the value metric for ATLAS.  

As far as we can tell from published or estimated numbers, the best 
condition $SS$ for other surveys at $m_s=20$ includes
$\sim$0.4~deg$^2$/sec for a single ASASSN unit ($g$ filter), $\sim$7 and $\sim$50~deg$^2$/sec for Catalina's Schmidt and 60" telescopes (unfiltered), $\sim$100~deg$^2$/sec for ZTF ($r$ filter), $\sim$700~deg$^2$/sec for each Pan-STARRS telescope (w filter), and $\sim$25,000~deg$^2$/sec for LSST ($r$ filter).  Use of filters causes a loss
of $SS$ but is offset by the scientific value of spectral information, and is a choice driven
by the goals of a survey system.
Of course different surveys are designed for different $m_s$, so ASASSN cannot do LSST's job nor LSST ASASSN's job, but we encourage the use of this methodology for design, optimization, and comparison of new survey capability.

\begin{figure}[h]
\begin{center}
\includegraphics[width=6in]{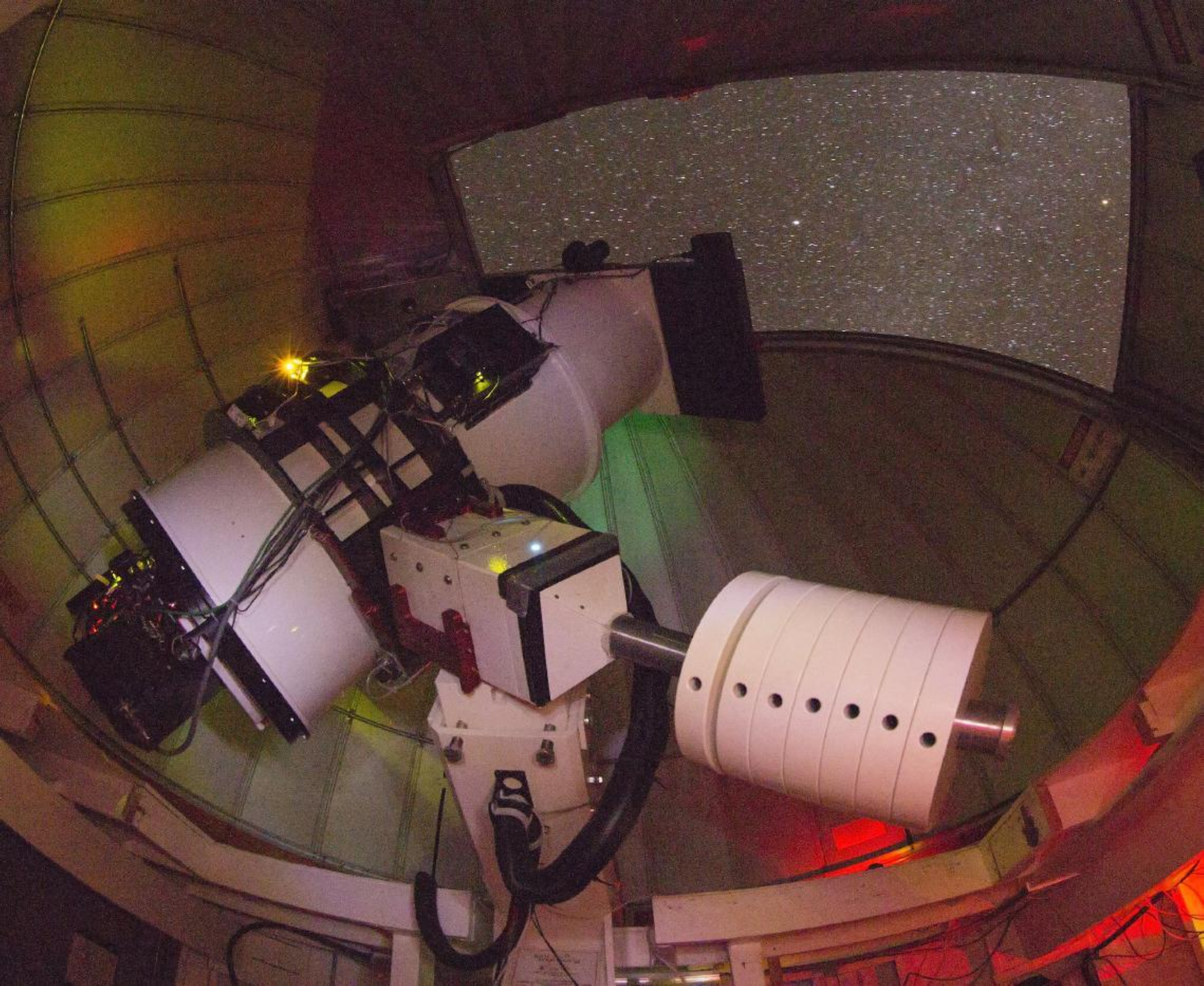}
\end{center}
\caption{The ATLAS unit on Haleakala from inside the dome.}
\label{fig:fisheye}
\end{figure}

Autonomous operation is another requirement for a cost efficient
survey, as well as enabling the low latency processing and discovery
essential for impending impacts.  For the NASA mission, ATLAS has built
a system that consists of summit operations, reduction pipeline, and a
science client that processes the output for moving objects.  The
summit operations automatically close, open, and observe when
possible, following an automatic schedule.  The reduction pipeline
calibrates the images, subtracts from them a static sky image, and
produces a table of detections of sources that have changed from the static sky.
The moving object science client waits until multiple observations
arrive for a given area on the sky and then links detections into
plausible asteroid tracklets.  Candidate unknown asteroids are screened by
a human for accuracy and are then posted to the Minor
Planet Center, who coordinates followup of unknown near-Earth asteroids.

Other science clients also tap the results from the  sky subtracted 
images. Using computer resources at Queen's University Belfast (QUB),  we 
search the same detection tables for 
stationary transients. These stationary transients are spatially matched against star, galaxy, 
active galactic nucleii, and QSO catalogues. The variable objects are filtered out leaving 
supernova candidates which are automatically reported publicly to the International Astronomical 
Union (IAU) Transient Name Server (TNS, Smith et al. {\em in prep.}). We also mirror all the ATLAS raw data at QUB as a safe, 
off-site backup.  ATLAS is designed and operated to be optimal for asteroid discovery,
but the sky survey synergistically contributes significant results in many other science areas.

\begin{itemize}
  \item{} Among regionally dangerous ($>30\mathrm{m}$) asteroids detected during very close approaches ($<0.01\mathrm{AU})$ to the Earth, ATLAS detects as many or more than any other asteroid survey, demonstrating its successful optimization as a `Last Alert' system for potential impactors.
  \item{} Up to the end of 2017, ATLAS reported 1175 candidate supernovae to the IAU Transient Name Server  \footnote{https://wis-tns.weizmann.ac.il}. Notable discoveries include detection of the shock break-out signature of SN2016gkg  \citep{2017ApJ...837L...2A} and the discovery of the unusual interacting type Ic supernova SN2017dio \citep{2017arXiv171200027K}. 
  \item {} Between 1 Jan 2016 and 18 Mar 2018, ATLAS has discovered the most spectroscopically classified transient objects (311, compared to ASASSN's 307), as reported in the TNS. This is enabling a host of ongoing science projects.  For example, through an alliance with the Public ESO Spectroscopic Survey for Transient Objects 
  \citep[PESSTO; ][]{2015A&A...579A..40S}, ATLAS provides young supernovae for the Foundation Supernova Survey \citep{2018MNRAS.475..193F}, working to create a definitive low redshift type Ia supernova sample to anchor cosmological analyses. 
  \item{} The large nightly ATLAS sky footprint has allowed searches for counterparts of gravitational wave sources from the LIGO - Virgo collaboration. During the first two observing runs ATLAS was a signatory to the agreement to share triggers. We searched for possible bright counterparts to binary black hole (BBH) mergers and discovered the afterglow of a gamma ray burst (GRB) before the high energy source was localized on the sky \citep{2017ApJ...850..149S}. This object, ATLAS17aeu, was discovered within the sky map of GW170104 \citep{2017PhRvL.118v1101A}, but is likely an unrelated GRB exploding 24hrs after the gravitational wave trigger. 
  This is only the third GRB afterglow detected independently of a high energy trigger \citep[the others discovered by][]{2013ApJ...769..130C,2015ApJ...803L..24C}.
  \item {}  A merging neutron star system produced the source GW170817 \citep{2017PhRvL.119p1101A,2017ApJ...848L..12A} and was accompanied by the discovery of an optical and near-infrared bright kilonova. It was discovered in NGC4993 at a distance of only 40 Mpc by several telescopes as soon as night fell in Chile \citep{2017Natur.551...64A,2017arXiv171005452C,2017ApJ...850L...1L,2017ApJ...848L..16S,2017ApJ...848L..27T,2017ApJ...848L..24V}. ATLAS had been continually observing  NGC4993  until  16  days  before  GW170817, and we showed it was not a variable source over the previous 601 days \cite[][see also Valenti et al. 2017]{2017Natur.551...75S}. 
  ATLAS will provide meaningful limits
  on the rate of kilonovae (irrespective of GW triggers) within 60\,Mpc \cite[][and Coughlin et al. in prep]{2018ApJ...852L...3S}.
  \item{} During its first two years ATLAS observed 140 million stars hundreds of times and
          has detected variability (pulsation, rotation, occultations, outbursts) in 5 million
          objects (Heinze et al. {\em in prep.}).  We will be releasing these lightcurves through the  Mikulski Archive for Space Telescopes (MAST). The ensuing 6 months has doubled the number of detections and increased the number of stars to 240 million, and there will be periodic data releases and updates.
  \item{} Asteroid characterization: color, rotation, volatile emission, and collisions are all
  measureable in the ATLAS lightcurves.
  \item{} ATLAS regularly detects satellites in geosynchronous orbit and beyond, and our multiple
  observations allow us to determine accurate 3D positions and velocities.
\end{itemize}

The NEA optimised survey strategy employed by the first two ATLAS units is equally good for transients and variables. We view the ATLAS unit as an inexpensive, reproducible system that could be deployed at sites judiciously separated  in latitude and longitude to give 24hr, all sky coverage to $m\sim20$ with a 1-day multi-exposure cadence. 
This paper is the first in a series describing the ATLAS hardware and software systems. It gives a broad overview of all the components that make the survey functional,  more specialised papers are in preparation giving details of the subsystems.

\section{Enclosure}

We considered a number of possible enclosures for ATLAS including
traditional Ash domes with an over-the-top shutter, clamshell designs
such as Astrohaven, and enclosures with roll-off roofs.  
We even designed ``ATLAS-in-a-can'', an ATLAS unit in a
commodity truck with a fold-off roof and a hole in the floor so the
mount could be lowered onto a solid pedestal.  As far as we know
``ATLAS-in-a-can'' would perform well, and with a very compliant shipping
truss to hold the ATLAS telescope and mount within the truck as well as
the truck's suspension, transportation would be simple and safe.

Concerns over wind buffeting, water leakage in severe storms, ambient
light and overall reliability led us to choose Ash domes as the ATLAS
enclosure.  Because we must operate autonomously, reliability is 
an extremely important consideration.
We do create unusual stress on the Ash dome by rotating every
$\sim$40~sec, which leads to bolts loosening, so threadlocking adhesive on most fasteners is a required upgrade from the nominal Ash construction.

The dome is a standard 16.5 foot diameter half-sphere over an 8 foot tall cylinder.  The pier is offset south from the center by 14 inches and is 41 inches tall and 30 inches diameter.  Steel rebar is epoxied into the slab to provide stiffening and support for the concrete pier.
We modified the Ash dome by adding I-beam mount points for a 500 pound rated chain hoist and by putting a Canarm 20-inch exhaust fan in
the wall which has motorized louvers that close when not in operation.
During daylight hours in good weather we open the dome slit by 6 inches for air intake and run the exhaust fan continuously. This keeps the interior of the dome close to ambient temperature. Without the fan, the sun beating down on the aluminum greatly increases the air temperature inside the dome, resulting in bad seeing during the early part of the night and unnecessary thermal stress on the equipment.

The Ash domes use a servo loop controlled stepper motor, and therefore
have precise acceleration and movement.  The absolute zero position is
set by a switch that is engaged at a particular position and the position thereafter is known by counts.  The domes use slip rings to
get power to the shutter motor, so have no limitations on rotation.
We have some concerns that our $\sim$900 dome moves each night may be
causing inordinate wear, but so far our monthly maintenance has
revealed no more than bolts vibrating loose and needing to be retightened.
The dome angular velocity is about 3.75~deg/sec and about 2~sec is
spent in acceleration, so the dome can move $\sim$25\deg\ during the
$\sim$9~sec of CCD readout and shutter overhead.

In addition to the normal electrical wiring required, we built a
``mezzanine loft'' and stairs that permit easy access to the telescope. The following equipment
supports the operations of mount, telescope, and camera:
\begin{itemize}
  \item{} Switch and fiber connections to our ``computer room''.
  \item{} Various ``low power'' industrial computers, one running
    Windows to interface to the DFM telescope (see Section\,\ref{sec:tel}), another running Linux with a dedicated ethernet to the camera controller.
  \item{} Various Raspberry-pi computers to provide IP network interfacing with individual devices (dome, Canon cameras, etc)
  \item{} ThermoFisher Accel 500 water chiller for CCD cooling.
  \item{} Puregas CDA-10 dehumidifier for mitigating moisture condensation on the camera window.
  \item{} A pair of webcams, a microphone, and an Ipod Touch to monitor the inside of the dome and communicate with personnel working at the summit.
  \item{} Keyboard and monitor for on-site manual control and system monitoring.
  \item{} A ``fail safe'' Raspberry-pi that uses a Hydreon rain sensor
    and monitors the electrical power to the dome.  If the power fails or the Hydreon reports
    rain or mist this computer closes the dome immediately.
  \item{} Various uninterruptible power supplies (UPS) to provide temporary power while the observatory safely shuts down automatically.
  \item{} A fisheye camera and meteorology box located nearby provides environmental telemetry (see Section\,\ref{sec:sys}).
\end{itemize}

\section{Mount}
\label{sec:mount}
We considered an equatorial mount to be  
essential, since an az-alt mount and image rotator adds significant risk of technical failure.  
Furthermore, 
the degraded performance near the
zenith for an az-alt telescope causes problems for an all-sky survey.

The ATLAS mount is  a German equatorial mount (GEM) built by APM Telescopes
of Saarbruecken Germany.  We considered a fork mount, but for a
telescope of this size a GEM is simpler, less expensive, and the advantage of cable routing for a fork mount
is not a significant factor at this scale.  The APM mount has 7 large counterweights
of about 35~kg apiece to counterbalance the mass of the
telescope.

This mount is very fast.  The slew velocity is 15~deg/sec,
and for moves smaller than 45 deg the time to slew and resume tracking
is $6.5\pm0.8$ sec, comfortably less than the CCD readout time.  A
meridian flip requires a rotation of $\sim$180\deg\ in both axes and
typically takes 25~sec.  Our scheduling software is mindful of the
cost of a meridian flip and minimizes them.

There are small issues with servo loop stability that occasionally
cause some image elongation.  Re-tuning the servo parameters
cures this, but we do not fully understand why it returns.  Winds higher than
40\,km\,h$^{-1}$ can buffet the large shutter at the top of the telescope depending on
dome position, elongating the images.
(The wind speed is measured by a Boltwood sensor; ATLAS is allowed to open for speeds below 30~kph and is required to close for speeds above 60~kph.)
The APM mount can track in both axes, and we have an elaborate and
accurate mount model.  Since our exposures are short, we do not
need to have perfect polar alignment.

\section{Telescope}
\label{sec:tel}
\begin{figure}
\begin{center}$
\begin{array}{ccc}
\includegraphics[width=3in]{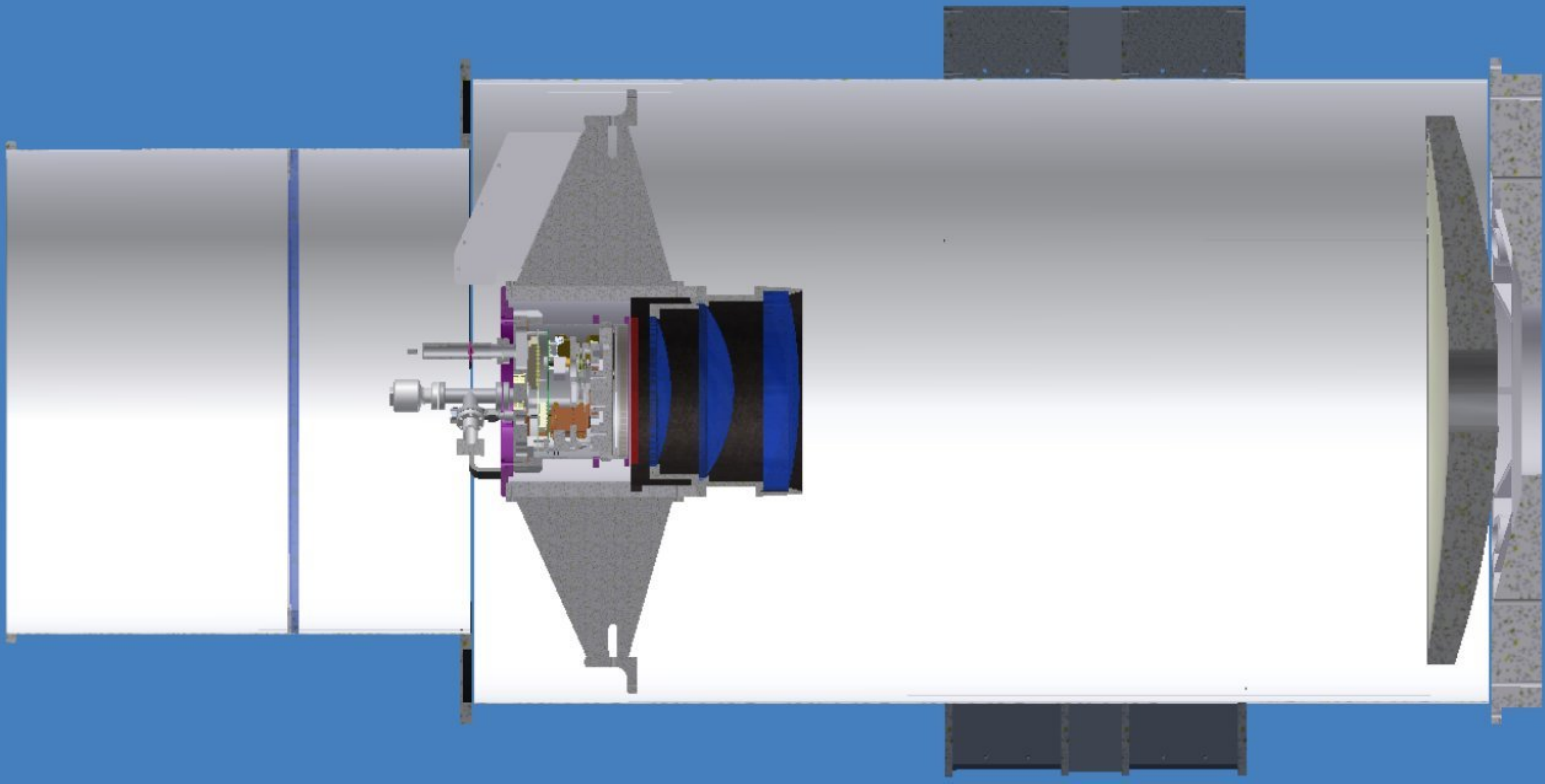}
\qquad
\includegraphics[width=3in]{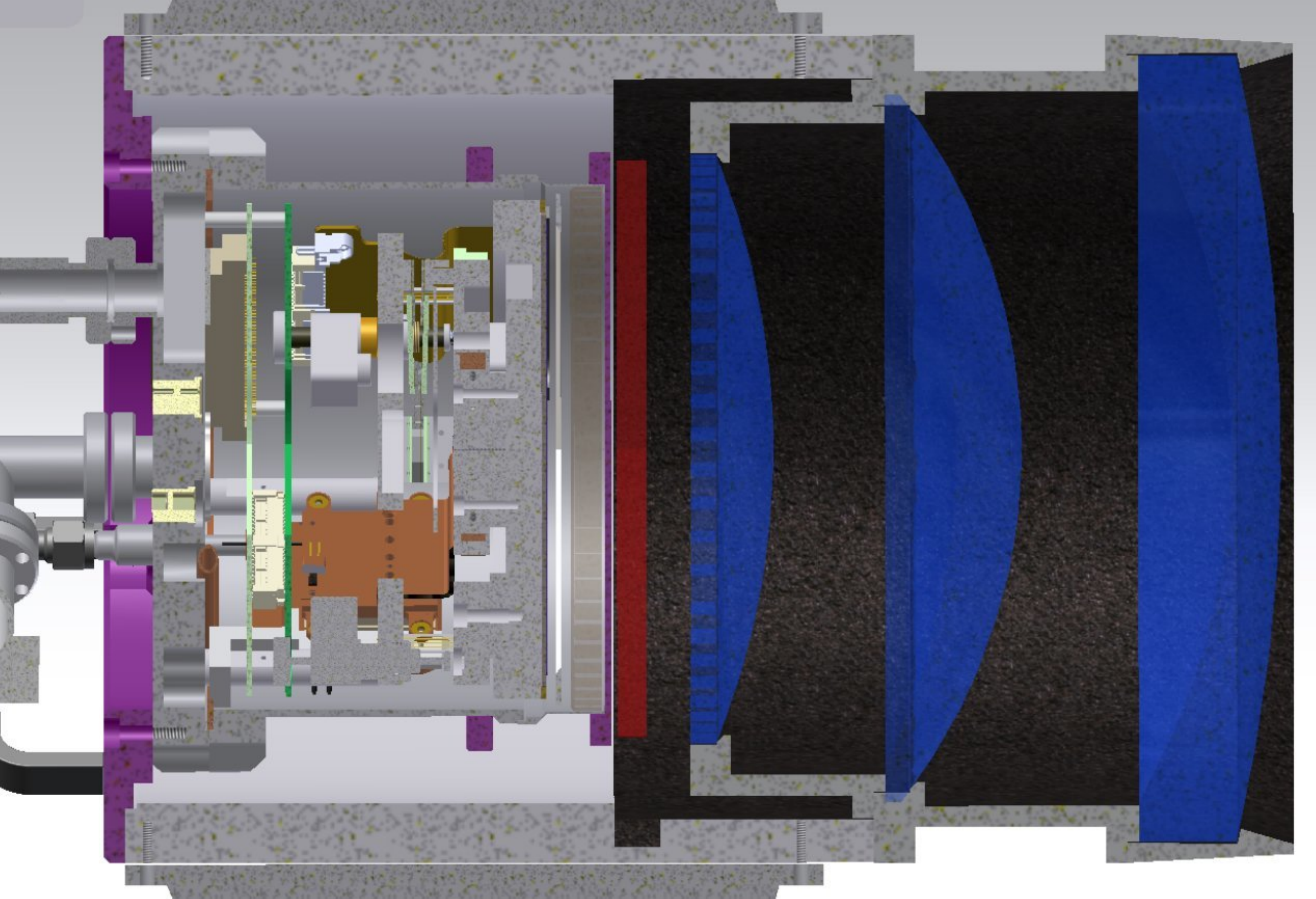}
\end{array}$
\end{center}
\caption{ Left: Light enters the Schmidt corrector of the DFM telescope (thin
  blue section on the left), passes the prime focus supported
  by a spider assembly, reflects off of the primary mirror on the
  right, returns through field correctors (three blue lenses), a
  filter (magenta), and a cryostat window (gray) before arriving
  at the detector. The overall length from shutter to back of mirror cell is 1.9~m, the focus unit adding another 0.4~m, and the diameter of the mirror cell is 0.8~m.  Right:  A detail of the field lens and camera assembly illustrates
  the close spacing of the third field lens, filter, cryostat window, and CCD.  The distance from Schmidt corrector to back of cryostat is 0.26~m, from there to the first field lens is 0.33~m, and the diameter of the field corrector and camera housing is 0.25~m.}
\label{fig:tel}
\end{figure}

Our telescopes were designed and built by DFM Engineering of Longmont, Colorado.  They are a
variant of a ``Wright Schmidt'', the optical train is a 0.5~m Schmidt
corrector, an 0.65~m spherical primary mirror, a three element field
corrector, a filter, the cryostat window, and the detector.
These are illustrated in Figure~\ref{fig:tel}.

The overall focal length is 1.0~m, for a system $f$-ratio of f/2.0.  The
optics perform well over a field diameter of about 7.5\deg, and are
designed to have modest chromatic aberration over the broad cyan bandpass ($c$, covering 
420-650~nm)  used by ATLAS, but images are distinctly sharper
in our redder survey bandpasss, called orange
($o$, 560-820~nm).  
The oversized primary mirror minimizes vignetting.  The field corrector assembly and camera housing
shown in Figure~\ref{fig:tel} has a maximum diameter of
250mm which determines the amount of central pupil obscuration.

The first telescope was installed on Haleakala in Jun 2015 and the
second in Feb 2017 on Mauna Loa.  The figure on the Schmidt correctors
was not perfect however, and the delivered image 
quality on the focal plane for the first
telescope was about 3.8 pixels FWHM (7\arcsec)
when initially installed. Efforts to improve
the second corrector at DFM were not successful so we initiated a
contract with Coherent Technologies (Tinsley) for a pair of corrector
lenses.  These were installed in May 2017 on both ATLAS units and the telescopes were
collimated.  The optics on Haleakala do an excellent job, producing
images slightly better than 2.0 pixels (3.5\arcsec), but there is some
residual astigmatism in the corrector on Mauna Loa, so the best images
are about 2.8 pixels (5\arcsec).  The astigmatism rotates with the
Schmidt corrector, but was not apparent in the Tinsley test results so
we are currently puzzled about its origin.

Collimation of the telescope is accomplished by adjusting push-pull screws
attached to the four ends of the spider assembly.  We
have written ray tracing software that calculates out-of-focus donuts
as a function of screw turns, so collimation proceeds by taking an
out-of-focus image, assembling a mosaic of donuts across the field of
view, judging from ray traces how many screw turns are required to
correct the donuts, and iterating.  We believe it prudent to keep
human judgement in this collimation loop.  Once collimated the
telescopes seem to hold their adjustment very well.

The focus is performed using an absolute encoder that seems very
accurate, and the telescope has an athermal design so there is
extremely little focus shift as a function of temperature.  The final
adjustment is the tip-tilt of the detector with respect to the focal
surface, and this is adjusted using the motors within the cryostat.
In-focus image elongation is a sensitive diagnostic of detector tilt.

DFM also designed and provided a full aperture shutter and a filter
changer.  The shutter uses bi-parting blades that are carefully balanced
to exert no force or torque on the telescope.  At this time the
shutter on Haleakala has operated nearly a million times and shows no
sign of wear or degradation.  The DFM filter changer comprises a
cassette that holds 8 filters in frames and lifts them to an insertion
mechanism that advances them into a slot between the last field corrector
lens and the camera.

ATLAS filters are 125mm square and 9mm thick.  We use broad band
filters for our normal asteroid search, a ``cyan'' ($c$) band from
420--650~nm, an ``orange'' ($o$) band from 560--820~nm, and a
``tomato'' ($t$) band from 560--975~nm intended to be differentially
sensitive to the silicate band of stony asteroids relative to $o$
band.  Haleakala normally switches between $c$ and $o$ during survey operations in a 
lunation, whereas Mauna Loa stays in $o$ or $t$.
Table~\ref{tab:coband} provides details of our primary filters as best we
currently know them.

We also have a set of filters in standard bandpasses, including one set of
Johnson/Cousins filters $B$, $V$, $R_c$, $I_c$,
and one set of  $g$, $r$, $i$, $z$ which are similar to 
SDSS and Pan-STARRS1 \citep{1996AJ....111.1748F,2012ApJ...745...42T}. 
ATLAS also has Skymapper-like ultra-violet filters 
$u$ and $v$ \citep{2011PASP..123..789B}
and two narrow band filters centered to trace $H\alpha$, and
[O\,{\sc iii}]. Discussions with the Skymapper team led us to 
adjust the center and widths of $u$, $v$, and $g$ to ensure
better delineation than those of \cite{2011PASP..123..789B}. 
The $o$, $c$, Johnson, and $H\alpha$ filters were
provided by Materion (Barr), and the rest by Asahi.  Details
found in Table~\ref{tab:coband} and Table~\ref{tab:filt} are calculated from manufacturer's curves
for the filters, AR coatings, 1.2 airmasses of atmosphere, 0.92
reflectivity of overcoated aluminum, and the measured detector QE.
The bandpasses have been adjusted for the ATLAS f/2 beam using an
effective index of $n=2$, but no in-situ measurements have been made.
Note that the $u$, $v$, and $z$ filters have low
transmission because the field corrector AR coatings are very
reflective outside of 380--850~nm. Should more ATLAS units be built
we intend to open up the IR and UV transmission of the optics.
Table\,\ref{tab:filt} summarizes the parameters of all the ATLAS bandpasses.
\begin{table}[htp]
\caption{ATLAS c and o Bandpasses}
\begin{center}
\begin{tabular}{rrrrrrrrrrrr}
\hline
\hline
$\lambda$&T$_c$&$\lambda$&T$_c$&$\lambda$&T$_c$&$\lambda$&T$_o$&$\lambda$&T$_o$&$\lambda$&T$_o$ \\
\hline
400 & 0.000 & 495 & 0.670 & 590 & 0.707 & 550 & 0.000 & 645 & 0.727 & 740 & 0.708 \\
405 & 0.000 & 500 & 0.678 & 595 & 0.709 & 555 & 0.001 & 650 & 0.727 & 745 & 0.700 \\
410 & 0.000 & 505 & 0.678 & 600 & 0.703 & 560 & 0.111 & 655 & 0.732 & 750 & 0.692 \\
415 & 0.007 & 510 & 0.688 & 605 & 0.708 & 565 & 0.642 & 660 & 0.735 & 755 & 0.682 \\
420 & 0.137 & 515 & 0.687 & 610 & 0.709 & 570 & 0.713 & 665 & 0.736 & 760 & 0.490 \\
425 & 0.509 & 520 & 0.699 & 615 & 0.709 & 575 & 0.707 & 670 & 0.737 & 765 & 0.491 \\
430 & 0.567 & 525 & 0.693 & 620 & 0.702 & 580 & 0.713 & 675 & 0.742 & 770 & 0.644 \\
435 & 0.564 & 530 & 0.689 & 625 & 0.697 & 585 & 0.717 & 680 & 0.747 & 775 & 0.644 \\
440 & 0.581 & 535 & 0.696 & 630 & 0.691 & 590 & 0.715 & 685 & 0.719 & 780 & 0.631 \\
445 & 0.589 & 540 & 0.703 & 635 & 0.716 & 595 & 0.717 & 690 & 0.681 & 785 & 0.625 \\
450 & 0.594 & 545 & 0.708 & 640 & 0.717 & 600 & 0.714 & 695 & 0.731 & 790 & 0.611 \\
455 & 0.599 & 550 & 0.715 & 645 & 0.699 & 605 & 0.715 & 700 & 0.737 & 795 & 0.600 \\
460 & 0.602 & 555 & 0.708 & 650 & 0.558 & 610 & 0.717 & 705 & 0.737 & 800 & 0.587 \\
465 & 0.603 & 560 & 0.710 & 655 & 0.042 & 615 & 0.713 & 710 & 0.735 & 805 & 0.570 \\
470 & 0.612 & 565 & 0.707 & 660 & 0.000 & 620 & 0.718 & 715 & 0.731 & 810 & 0.562 \\
475 & 0.627 & 570 & 0.704 & 665 & 0.000 & 625 & 0.723 & 720 & 0.705 & 815 & 0.531 \\
480 & 0.647 & 575 & 0.697 & 670 & 0.000 & 630 & 0.718 & 725 & 0.703 & 820 & 0.365 \\
485 & 0.657 & 580 & 0.710 & 675 & 0.000 & 635 & 0.726 & 730 & 0.702 & 825 & 0.021 \\
490 & 0.663 & 585 & 0.713 & 680 & 0.000 & 640 & 0.721 & 735 & 0.710 & 830 & 0.000 \\
\hline
\end{tabular}
\end{center}
\tablecomments{The first six columns give wavelength [nm] and transmission of the
  ATLAS system in $c$ band, the last six columns in $o$ band ($o1$).
  The transmission includes 1.2 airmasses of atmosphere, 
  mirror reflection and AR coatings, filter transmission calculated
  for an f/2 beam, and detector QE.}
\label{tab:coband}
\end{table}%

\begin{table}[htp]
\caption{ATLAS Bandpass Parameters}
\begin{center}
\begin{tabular}{lcrrrrlcrrrr}
\hline
\hline
  Filter & $\langle T\rangle$ & $\lambda_{eff}$ & $d\ln\nu$ & $\lambda_B$ & $\lambda_R$ &
  Filter & $\langle T\rangle$ & $\lambda_{eff}$ & $d\ln\nu$ & $\lambda_B$ & $\lambda_R$ \\\hline
  $c $   &   0.669 &  533 &  0.290 &  423 &  651 &     $u $   &        0.032 &  353 &  0.002 &  344 &  362 \\
  $o(1)$ &   0.684 &  679 &  0.261 &  563 &  818 &     $v $   &       0.343 &  394 &  0.027 &  379 &  407 \\
  $o(2)$ &   0.680 &  678 &  0.262 &  561 &  819 &     $g $   &       0.638 &  484 &  0.175 &  420 &  549 \\
  $t $   &   0.641 &  710 &  0.320 &  558 &  876 &     $r $   &       0.713 &  619 &  0.166 &  552 &  691 \\
  $B $   &   0.545 &  445 &  0.125 &  394 &  494 &     $i $   &       0.639 &  751 &  0.112 &  692 &  818 \\
  $V $   &   0.621 &  548 &  0.095 &  509 &  589 &     $z $   &       0.333 &  892 &  0.054 &  826 &  940 \\
  $R_c $   &   0.665 &  652 &  0.162 &  577 &  732 &     $H\alpha $ &   0.592 &  656 &  0.009 &  652 &  660 \\
  $I_c $   &   0.554 &  794 &  0.107 &  728 &  876 &     [O\,{\sc III}]&0.592 &  505 &  0.014 &  500 &  509 \\
\hline
\end{tabular}
\end{center}
\tablecomments{The columns are the filter, mean transmission of 
  the system including 1.2 airmasses of atmosphere over the blue and
  red limits, filter ``pivot'' wavelength [nm] described
  by \cite{2012PASP..124..140B}
  ($\int\lambda T(\nu)d\ln\nu/\langle T\rangle$), bandpass width in
  log wavelength times transmission (canonical 0.2 times $\epsilon$ in Equation~\ref{eq:etendue}), and bandpass blue and
  red wavelengths [nm] obtained from a least-squares fit of a square
  bandpass. 
  $o(1)$ is installed on Haleakala and $o(2)$ on Mauna Loa.  }
\label{tab:filt}
\end{table}%

Approximate conversions between ATLAS primary filters and Pan-STARRS $g$, $r$, and $i$ filters
for stellar spectral energy distributions are given by Equation~\ref{eq:filt}.

\begin{align}
  c     & \nonumber \sim 0.49\, g + 0.51\, r \qquad
  o \sim 0.55\, r + 0.45\, i \\
  g     & \nonumber \sim  1.67\, c - 0.67\, o \qquad
  r \sim  0.35\, c + 0.65\, o \qquad
  i \sim -0.39\, c + 1.39\, o \\
  (c-o) & \sim 0.73\, (g{-}r) \sim~ 0.47\, (g{-}i)
\label{eq:filt}
\end{align}

\section{Camera}

The ATLAS camera was required to satisfy a number of requirements that could not
be met by any commercial product, therefore we designed and built it in house.
These requirements we imposed included:
\begin{itemize}
  \item{} Fill as much of the 130~mm diameter optical field of view as
    possible, but sample the expected PSF of 7\,$\mu$m RMS with pixels no
    larger than 10\,$\mu$m. 
  \item{} Read out in less than 10~sec so that the duty cycle for a
    30~sec exposure is no worse than 0.75, with a read noise of no
    more than 10~e$^-$.
  \item{} Camera diameter can be no larger than 200~mm, length no
    longer than 200~mm, mass no more than 7~kg.
  \item{} The cryostat window is no more than 10~mm thick and distance to detector
    surface may be as little as 6~mm.
  \item{} The detector must be colder than $-50$~\degc\ for dark
    current to be negligible compared to the sky background.
  \item{} There must be a means to remotely tip and tilt the detector
    to align with the f/2 focal surface.
  \item{} All connections to the camera must pass along a 3/4-inch
    channel on a spider vane, and the detector controller may be a distant as 0.5~m.
  \item{} The cryostat must maintain a vacuum of 
    1~mtorr or less for at least a year (preferably much longer) because
    of the difficulty in extracting the camera from the center of the
    telescope.
\end{itemize}

After a competitive procurement we selected STA (Semiconductor Technology Associates of San Clemente 
Califronia) as the vendor for the 
CCDs, and chose their STA-1600 as the ATLAS detector.  This is a
monolithic CCD with 10560$\times$10560 9\,$\mu$m pixels, thinned,
passivated, and AR coated by the Imaging Technology Laboratories of the University of Arizona.  STA also provided the controllers
and cables.  We collaborated closely with STA, both on the mechanical
and electrical interfaces of fitting the CCD inside the cryostat as
well as tasking STA to build a custom board for ATLAS auxiliary
functions such as temperature and pressure monitoring, thermoelectric
cooling, and piezoelectric motor operation.

The cryostat consists of a 6.5-inch diameter ``bell jar'' that has a 3/8-inch
fused silica window brazed on one end, and a standard CF-8 flange on
the other end which is the solid base plate on which the internals are
mounted.  Outside the baseplate are a metal seal vacuum valve, an MKS
micro-Pirani vacuum gauge, a Modion ion pump, a warm zeolite getter,
and connectors for cooling water.
Inside the cryostat are a pair of water-fed heat exchangers,
a pair of two stage thermoelectric coolers (TEC), and a pair of pyrolytic
graphite cold straps to a cold plate that carries the CCD.  The heat
exchangers have bistable thermal switches to disconnect TEC power at
+50~\degc\ if water flow is interrupted.   The cold
plate is mounted on top of a flexure and is pulled down by a set of
springs and pushed up by a trio of ``picomotors'', vacuum rated units
that use piezoelectric slabs to turn a fine pitch screw with nanometer
precision.  They have an enormous travel, 12~mm in our case, and are
electrically inert when not in use.  The position of the cold plate is
monitored using linear Hall sensors that measure the radial field near
the center line of a cylindrical magnet.  Although we close the
position loop by observing stars, we calibrate these Hall sensors in
the lab using a microscope to observe the cold detector through the
window, and we achieve absolute accuracies of about 1\,$\mu$m

A pair of printed circuit boards from STA are mounted within the
cryostat and are cooled by the heat exchangers. These buffer drive signals to
the CCD and convert the CCD output to true differential signals that
pass along the cable to the controller.  This provides us with
excellent noise immunity and we see no interference of any sort, even
when the ion pump or picomotors are active.

\begin{figure}
\begin{center}$
\begin{array}{ccc}
\includegraphics[height=2.5in]{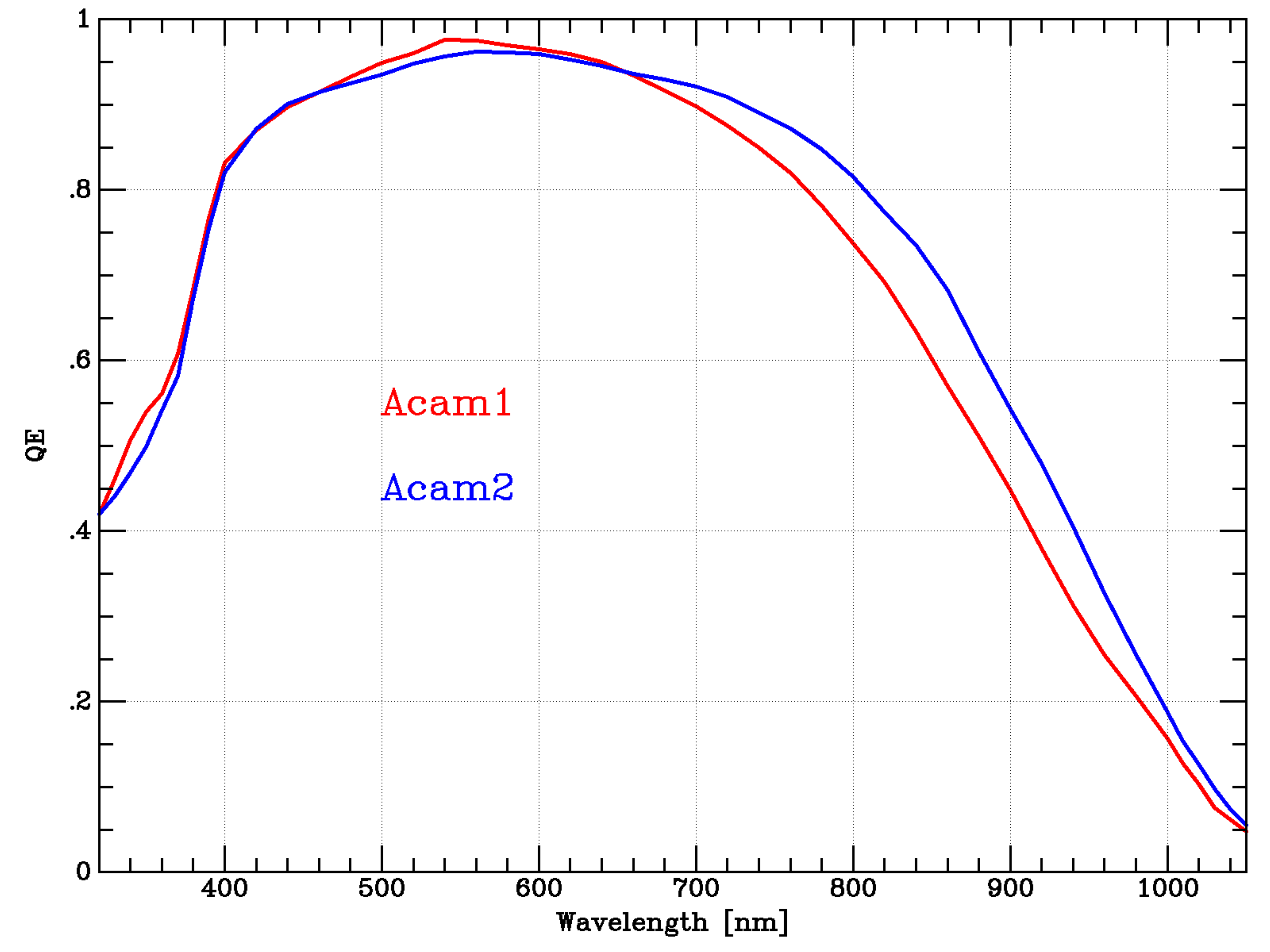}
\qquad
\includegraphics[height=2.5in]{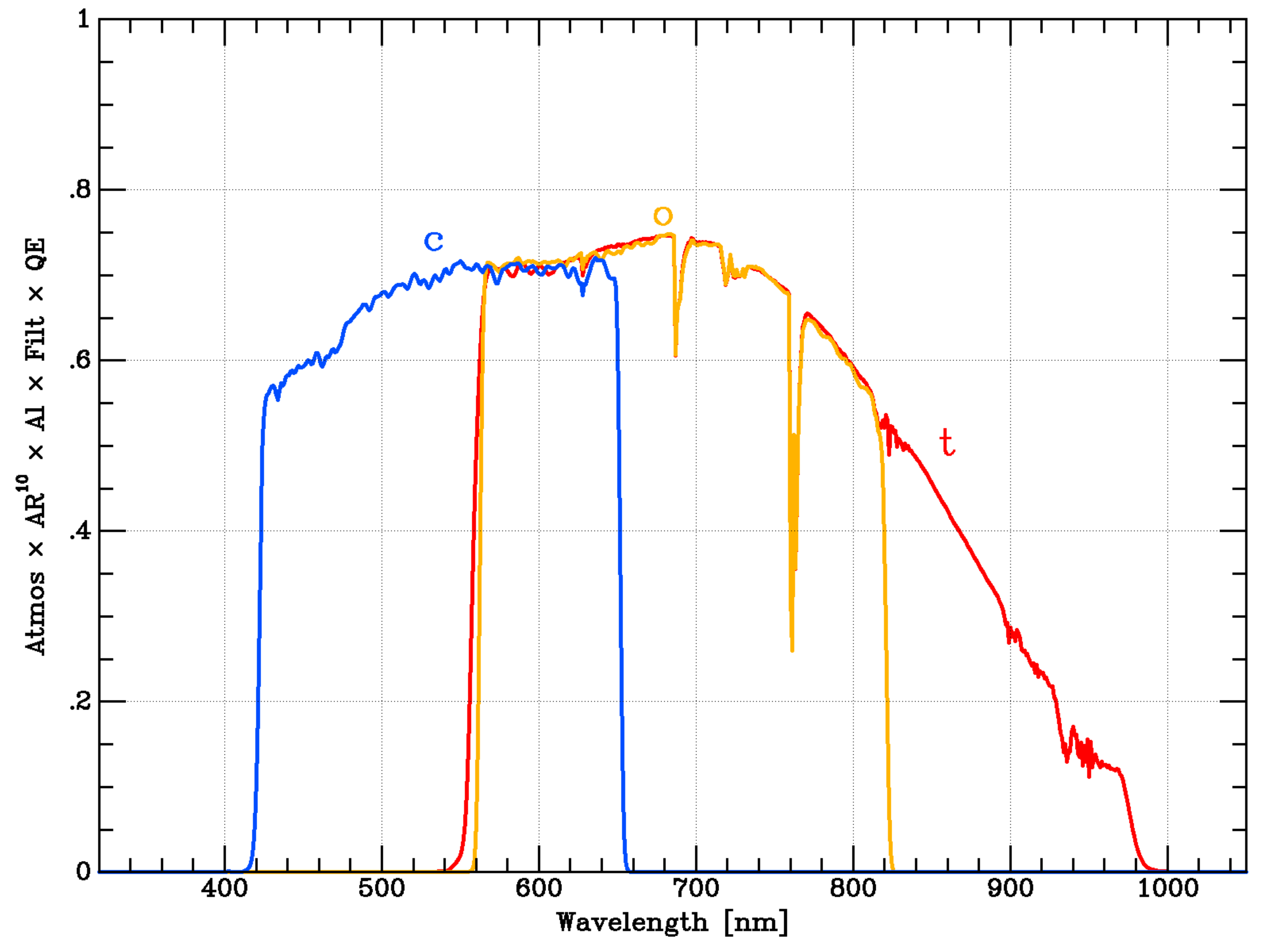}
\end{array}$
\end{center}
\caption{Left: The quantum efficiency measured at $-50$~\degc\ relative to a Hamamatsu
  photodiode with NIST-traceable calibration for out two 
  CCDs. Right: The calculated total throughput from multiplying
  the traces of our primary $c$ (blue),
  $o$ (orange), and $t$ (red) filters (converted to f/2), 1.2 airmasses of extinction, AR
  coatings, and a factor of 0.92 for the enhanced coating on the
  mirror.}
\label{fig:qe}
\end{figure}

The CCDs are set to a gain of about 2~e$^-$/ADU, and they have a full
well in excess of 80,000~$e^-$.  We read out at 1~MHz through 16
amplifiers so the total read time is about 9~sec.  With our normal
30~sec exposure this gives a shutter-open duty cycle of about
75\%.  The read noise is about 11~e$^-$ and in a 30 second exposure
the typical moonless sky background is about 300~e$^-$ in $c$ band and
350~e$^-$ in $o$ band, so the read noise degrades our SNR by about
16-18\%.  
The quantum efficiency and total throughput through our two primary
filters are illustrated in Figure~\ref{fig:qe}.

The dark current in our STA1600 CCDs has quite a bit of pattern to it,
particularly between 8 horizontal bands.  The average is about
0.8~e$^-$/pix/sec at a CCD temperature of $-50$~\degc, rising a factor of 2
every 5~K.  The thermoelectric coolers keep the CCDs at a temperature
of about $-53$~\degc, so dark current is not a significant contributor to
noise in our wide 250~nm $c$ and $o$ filters, but is a concern for
our 10~nm wide narrow band filters or our UV filters.  The water to
the heat exchangers in the cryostat is cooled to $+6$~\degc\ by a ThermoFisher
Accel 500 chiller, and the detectors cool by 2~K for each 3~K decrease in
water temperature, so it is feasible to lower the dark current to
about 0.2~e$^-$/pix/sec when required.

Detector flatness is a concern in our fast, f/2 beam.
Scans of the surfaces of Acam1 (HKO) and Acam2 (MLO) through the cryostat window
when cold reveal an RMS deviation from flatness of 4um for Acam1 and 9um
for Acam2 and a peak-to-peak deviation of 18um and 36um.
Although Acam2 is a factor of 2 worse than Acam1, the maximum blur
circle is only 9$\mu$m (1.8\arcsec) and the RMS is half of that.


These CCDs are cosmetically quite good: Acam1 has a region with bad
charge transfer efficiency of about 0.2\% of the area and Acam2 has a few blocked columns.
The thinning leaves some artifacts but these flatten quite well; about
1\% of the area is lost to the thinning border.  Both of these CCDs
have peculiar, low level flattening artifacts that are particularly
evident in some of the horizontal sections.  Acam1 has little
``dipoles'', where charge from one pixel is borrowed by the one
immediately below it, and Acam2 has a ``bamboo forest'', where
adjacent columns seem to exchange charge in a wavey pattern.  It is
believed that these are the result of CCD manufacturing masks with
insufficient resolution, and more recent CCDs are better.  These artifacts
flatten quite well, so for our purposes they are not
important.

Fast readout confers other problems such as bias levels that have a
``plaid'' pattern and must be corrected by using both serial and
parallel overclocks, and cross-talk.  Our $16\times16$ cross-talk
matrix shows values ranging between $2\times10^{-4}$ to
$3\times10^{-5}$, with prominent correlations between amplifiers and
video boards.  The cross-talk diminishes dramatically with slower
clocking and we had some success in reducing it by carefully selecting
pedestal and signal samples.  We mark our output mask for each image
with a bit indicating pixels that might be under the influence of
cross-talk from some bright star, and we subtract a fraction of the
image from itself according to the cross-talk matrix, but
the cross-talk is not a serious issue.

Without any elastomer seals, and thanks to the ion pumps the cryostats
hold vacuum very well; the cryostat on HKO was last pumped three years ago
and shows no sign of leakage.

\section{Computers and software}

Because we want our system to be able to run autonomously even when
there is no internet connection to the observatory, we maintain a rack
of computers on each summit.  
This rack is in a ``computer room'' (in
fact just a reasonably dry and safe place separate from the dome), and
carries an ethernet switch and six 1U ``Supermicro'' server computers.  One is
dedicated to be our gateway, one is an ``admin'' computer that is
devoted to running the camera, and four are general purpose compute
nodes.  These computers each have about 12~TB of RAID1 disk, 12-16 cores (24-32 threads),
and 128~GB of memory, so they are very capable and each can save
approximately a month's worth of observation.  They provide
substantial redundancy in case of failure.

A normal night of observation produces approximately 900 images from
each of the main cameras, the auxiliary 35mm cameras, and the fisheye
35mm cameras, for a total of about 150~GB of raw, compressed data per
night. 

Given the good bandwidth to the observatories, our minimum latency
strategy has been to perform the first reduction steps of computation
at the observatory and the image subtraction and science processing at our
``base cluster''.  This strategy doubles the volume of data that
needs to be stored at the observatory and copied, but it allows us to
operate when bandwidth is poor or non-existent, and it avoids a future
bottleneck if ATLAS units were to proliferate.  We  have a hard requirement
of a latency of
less than an hour from shutter close to final results, because that is
the interval between the first and last observations of each field on
a given night (we observe 4 or 5 times at each position across a 1 hour 
period).  If bandwidth becomes an issue it is simple to carry out all the reductions at
the observatory and copy only the final detection tables in real time,
using the daylight hours to copy the images.

Our ``base cluster'' in Honolulu
currently consists of 16 of these 1U Supermicro
compute nodes and 5 4U storage computers.  The storage computers carry
24 disks apiece with hardware RAID6 and each provides 120~TB (160~TB
with more modern 8~GB disks) of storage.  
A 1U node costs about \$3,500, and a 4U storage computer
costs about \$12,000.  This is far less expensive than
cloud storage and computation, and although we foresee eventually
moving our data and processing to the cloud, it is not cost effective
to do it at present.

We have adopted a philosophy of ``less is better'' as regards
scripting languages, so we avoid high-level languages such Perl, Python, Java, or other
variants with complex dependencies so that we may provide a simplified computing environment.  We have
instead restricted our diversity of languages to C, bash and the usual
Unix tools, and Google's {\tt go} language\footnote{\tt https://golang.org/} for the telescope control system.  This creates a
significant benefit in terms of stability, computation efficiency, and
most critically comprehensibility between the various software
developers and users.  By judicious creation of ``Unix tools'',
meaning programs that are designed to be used like any other Unix
utility and have a man page, we have not only managed to get efficient
code written efficiently, we have also created a system that is more
portable, agile, and less complex than the usual GUI-oriented, scripting
language rat's nest.

Our software systems are described in detail elsewhere (Denneau et al. {\em in prep.}), but broadly
speaking the components are:
\begin{itemize}
  \item{} A telescope control system (TCS) deployed as a collection of lightweight executables written {\tt go} that share system state through a Redis database.
  \item{} A scheduler that creates the desired pointings for each
    night and executes them.
  \item{} A reduction pipeline that is responsible for converting raw
    camera files into flattened, calibrated images.
  \item{} An image subtraction pipeline that matches all-sky reference images (``wallpaper'') 
    to each image, subtracts it, and finds all the remaining sources. This employs 
    a modified version of \texttt{hotpants} for  image subtraction \citep{2015ascl.soft04004B}.
  \item{} ``Science clients'', specialized pieces of software that
    receive final images or tables and perform object selection and scientific analysis.
  \item{} A post-processing pipeline that executes tasks that are
    important but not time critical e.g. final photometry of all stars in a 
    reduced image.
\end{itemize}
Our primary science client is an adapted version of the Pan-STARRS Moving Object Pipeline System
(MOPS; \cite{denneau2013}), that links detections from different images into a plausible
moving object and reports observations to the IAU Minor Planet Center (MPC). 
Typical execution times for a 10k observation to be processed from telescope pointing to MPC reporting 
are shown in Table\,\ref{tab:proctime}. 
The photometry and detection programs
\texttt{dophot} 
\citep{1993PASP..105.1342S,2012AJ....143...70A}
and \texttt{tphot},
(described in a future paper, performance reported by \cite{2013PASP..125..456S}) 
are multithreaded, so the elapsed time is less than the CPU time.  
The total processing time from the moment the telescope settles on tracking the field center
and the shutter opens 
to having an object catalog with calibrated photometry and astrometry is 
approximately 40 minutes 
of CPU time and 25 minutes of real time.  The code could be optimized
for somewhat better performance, but right now we are focused on higher priority
development such as better science processing, better scheduling, and
documentation.
The deep
photometry of all stars in the reduced image is an example of
post-processing that is not required for our time critical science
client, and this runs on some of the redundant summit computers.

A second science client is the ATLAS Transient Server that runs on 
a computer cluster at Queen's University and links individual stationary detections 
into objects and reports supernova candidates to the IAU Transient Name Server
(Smith et al, in prep.). It runs after the final detection table is created and
uses the same input files as the moving object pipeline.

\begin{table}[htp]
\caption{ATLAS Processing time}
\begin{center}
\begin{tabular}{rrl}
\hline
\hline
  CPU & Elapsed & Stage \\
\hline
  40  & 40 & take exposure, save to disk as a raw image \\
  40  & 80 & flatten image \\
  500 & 250 & measure the brightest $\sim$60,000 stars (dophot) \\
  20  & 270 & find initial astrometric solution \citep{2010AJ....139.1782L}\\
  10  & 280 & determine final astrometric and photometric solution \\
  20  & 300 & perform cloud detection and correction \\
  10  & 310 & calculate auxiliary metadata, compress and save image \\
  \dotfill & \dotfill & \dotfill \\
  250 & 560 & produce wallpaper template matching image \\
  600 & 1160 & subtract wallpaper from image (hotpants) \\
  750 & 1410 & detect sources in difference image to $3\sigma$, trim to $5\sigma$ (tphot) \\
  180 & 1590 & classify sources, write final detection table \\
  120 & 1710 & run primary science client MOPS to detect moving objects \\
\hline
  900 & 610 & measure $\sim10^{5}$ to $\sim2\times10^{6}$ stars (depending on galactic latitude) to $5\sigma$ (dophot)\\
\hline
\end{tabular}
\end{center}
\tablecomments{The typical CPU processing time (sec), cumulative elapsed wall clock time (sec),
  and processing stage for each 10k image are
  listed. Stages above the dotted line occur on the summit computers and steps below 
  are run at the ``base cluster" in Honolulu. Below the solid line are the ``post-processing'' stages that are
  run separately so as not to increase the latency of primary
  difference detections.}
\label{tab:proctime}
\end{table}%

\section{System Features}
\label{sec:sys}

We have implemented a number of features to support autonomous
operations and to ensure high quality data.  For meteorological
information we have a small ``metfish'' that consists of a Boltwood
CloudSensor system, a Garmin GPS, and a Canon 10mm f/4 fisheye lens on a Canon
5DIII body.  This ``metfish'' is a watertight box that uses an AR-coated
glass dome intended for underwater diving to protect the lens.  The
Boltwood system reports temperature, wind speed, humidity, rain, and
uses a thermal IR pixel to examine the sky for clouds on a 1 second
cadence.  The fisheye camera in our ``metfish'' is similar to commercial 
cameras\footnote{\tt  http://www.alcor-system.com/new/index.html},
but with additional functionality and software.  A ``metfish'' costs
about \$10,000. We use the GPS to synchronize a Stratum~1 time server, and  
Network Time Protocol (NTP) brings
the rest of our computers to absolute time with an accuracy of a few
microseconds, regardless of network connectivity to the external world.

We use the geosynchronous satellite Galaxy-15 to calibrate our shutter
latency.  Because Galaxy-15 is part of the GPS constellation, its
position is tracked at the centimeter level by JPL and published by
the National Satellite Test Bed web site\footnote{\tt http://www.nstb.tc.faa.gov/rt\_waassatellitestatus.htm}.  We find that the
time between initiation of a shutter movement command and when the
shutter blade is half way across the aperture is $0.281\pm0.017$~sec
(this uncertainty arises for each Galaxy-15 observation; 
the repeatability suggests that the average delay is known better).
Obviously our absolute exposure accuracy is limited to a few
milliseconds, regardless of computer clocks that are accurate to the
nearest microsecond.

The fisheye camera takes images every 5 minutes during the day, but at
night switches to 32 second exposures on a 40 second cadence.  
The fisheye cameras at HKO and MLO are staggered by 20 seconds
so that one shutter is always open in Hawaii at night.  These
images are treated as scientific data and subjected to the same
rigorous flattening, astrometric, and photometric calibration.  We
achieve astrometric residuals of about 0.1 pixel, primarily limited by
the undersampled PSF and co-adding the color pixels into monochrome
super-pixels, and we can achieve photometric accuracy of 0.02 mag
for suitably bright stars.

The 5$\sigma$ limiting magnitude for a single fisheye image is about
$m=7$ (depending on declination and the degree of trailing), but we
have found that the sensitivity improves as $N^{1/2}$ for $N$ much larger
than 1000.  Thus a source which has average magnitude over 40 seconds
of $m=7$ is detectable, likewise $m=9.5$ over an hour or $m=12$ over
an entire night.  The fisheye pixels are about 4.3~arcmin, so very
faint sources are confused in stacked images, but image subtraction is
very effective at removing the non-varying background.

\begin{figure}[H]
\begin{center}$
\begin{array}{ccc}
\includegraphics[height=2.5in]{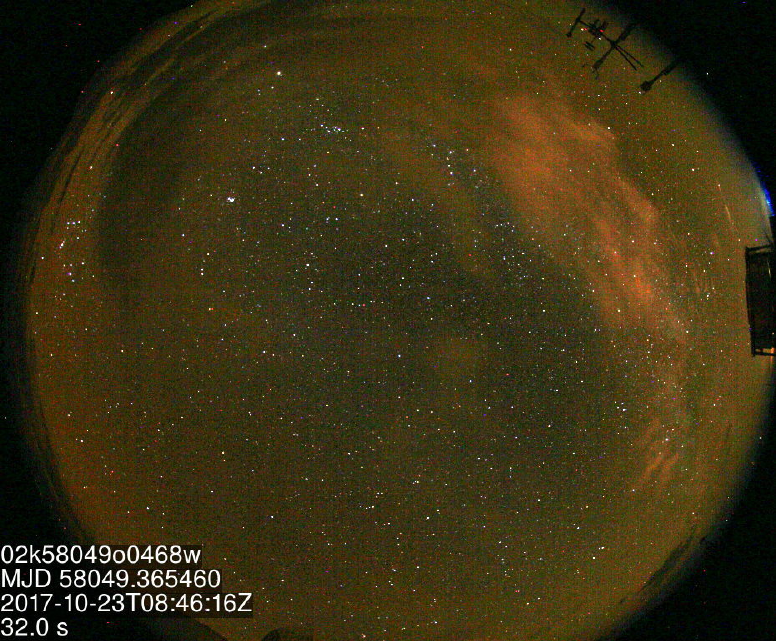}
\qquad
\includegraphics[height=2.5in]{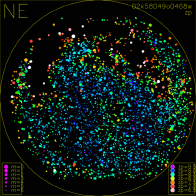}
\end{array}$
\end{center}
\caption{The left image shows a fisheye view of a typical partially
  cloudy night.  Clouds are evident, but their visibility depends on
  the illumination (sodium light from Honolulu) and their opacity is
  impossible to quantify.  The right image shows the zero point derived
  for 5,027 stars in the field, with transparency indicated by point color.  
  The clouds are obvious, and it is
  also possible to quantify the effects of faint clouds and
  extinction.}
\label{fig:fish}
\end{figure}

Each fisheye image provides us with extinction measurements
for up to 10,000 stars and it is easy to quantify where clouds are,
how opaque they are, and how they are moving.  We intend to develop
a scheduler that is responsive to this information.  
Figure~\ref{fig:fish} illustrates the utility of the fisheye on a
night when partial clouds interrupted otherwise clear sky.
The fisheye
images are visible from our public website\footnote{\tt http://www.fallingstar.com}.

Each dome is equipped with a Hydreon rain sensor, and each dome is powered
by a small 1350~VA UPS that has enough power to close the shutter
on a power failure.  A Raspberry-pi monitors the Hydreon and the AC
power and closes the dome when necessary regardless of what the rest
of the system is doing.

Finally, we mount a Canon 5DIII and Canon 135mm f/2 lens on
the telescope and take images synchronized with the main science
camera.  The field of view is 15\deg$\times$10\deg, easily
encompassing our main field of view, and the 5$\sigma$ limiting
magnitude is about $m\sim14$
in a 25 sec exposure.  These data are processed exactly
like any other, and therefore the combination of the three optical 
instruments allows us to monitor the sky over $0<m<20$.
A summary of the ATLAS camera system is in Table\,\ref{tab:camsum}. 

\begin{table}[]
\centering
\caption{Summary of the ATLAS system cameras}
\begin{tabular}{lllllll}
\hline
\hline
Camera                 &  Format             & Pixel size   &  Pixel Scale  &  FOV      & Dynamic range & Exp. Time\\
   &    (pixels)         & ($\mu$m)     &  (arcsec)     & (degrees) &  (sat. to 5$\sigma$) & (sec)\\
\hline
Acam                   & 10560$\times$10560  & 9            & 1.86 &   5.375$\times$5.375 & 11--19.5 & 30  \\
Canon\,135mm\,f/2        & 2880$\times$1920  & 12.6         & 19.4 &   15.5$\times$10.3   & 4--14    & 25  \\
Canon\,10mm\,f/4\,fisheye & 2475$\times$1920 & 12.6         & 255  &   180$\times$150     & 0--7     & 32  \\
\hline
\end{tabular}
\label{tab:camsum}
\tablecomments{The exposure time column refers to the typical exposure time during survey operations.
The 35mm camera's pixel size refers to summed RGGB Bayer sub-pixels, the color information is
kept in raw frames but not otherwise processed.}
\end{table}

\section{Performance Results}

Excluding the borders of the CCDs, our net field of view on the ATLAS Acam 
cameras is 
5.375\deg\ square for a total of 28.9~deg$^2$, and the
900 exposures taken during a night cover 26,000~deg$^2$. The 
declination range $-45\deg<\delta<+90\deg$ encompasses
85\% of the sky (35,065~deg$^2$) and
25\% of the sky lies within 60\deg\ of the Sun, which is essentially
unobservable. This leaves about 24,500~deg$^2$ of sky accessible on a given night.
A single ATLAS unit could therefore cover the entire accessible sky in
one night with a single 30 second exposure at each pointing. 

Our mission for NASA requires us to distinguish moving objects from
stationary transients, to provide a meaningful trajectory for moving
objects, and to have minimal false alarms.  We therefore observe each
field four times in a given night and reduce our Dec coverage for each
unit so as to cover 1/4 of the visible sky.  With two ATLAS units, our
four exposure coverage is 1/2 of the visible sky each night. 
We therefore cover the entire accessible sky with a cadence of 2 days, with 
four exposures (over a 1 hour interval) reaching 
$o\sim19.5$ in each individual frames 
when the sky is dark and seeing good.  For stationary transients the
co-adds of a night's exposures reaches $o\sim20.2$.

\begin{figure}[H]
\begin{center}
\includegraphics[width=4in]{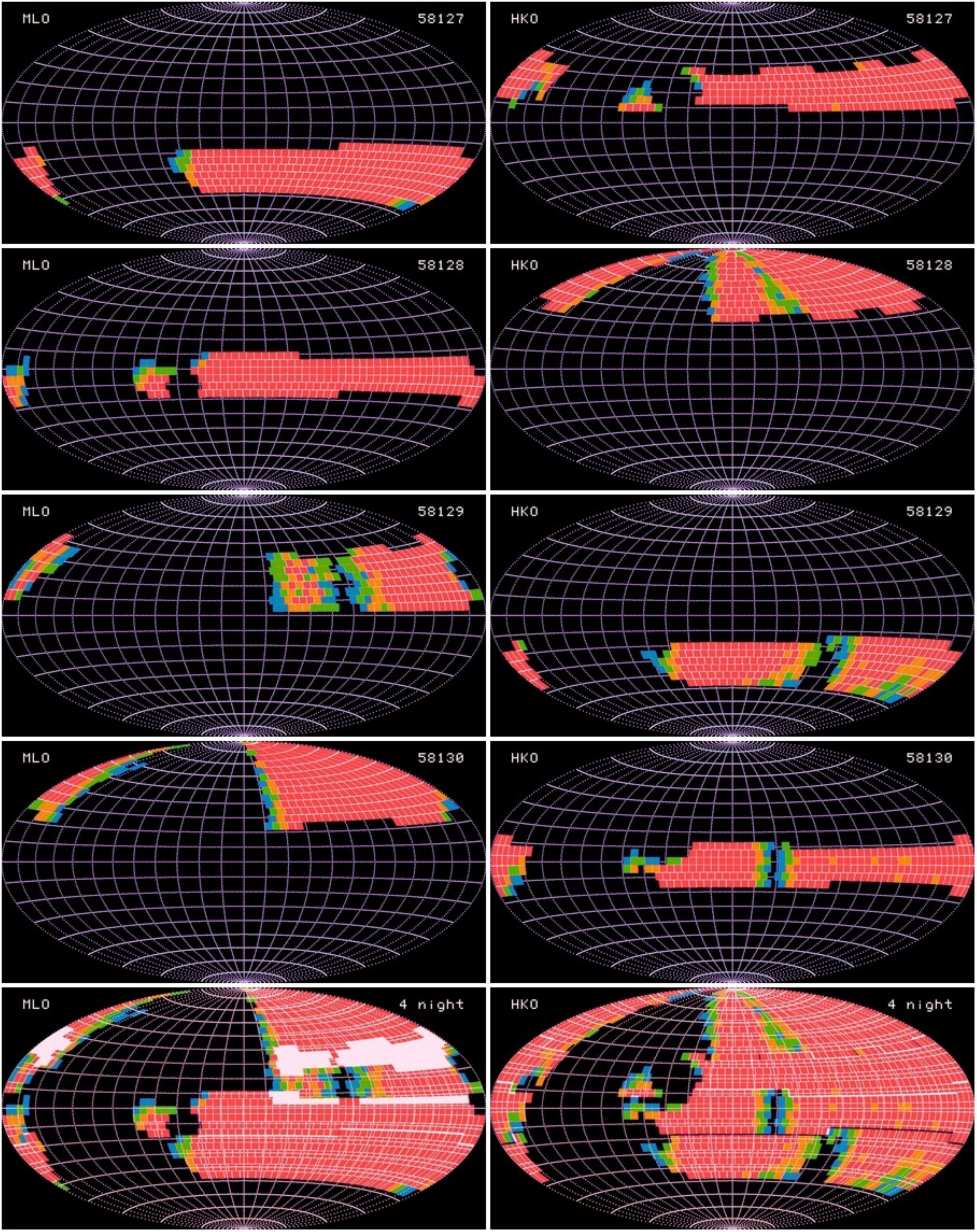}
\end{center}
\caption{Recent ATLAS coverage of the sky over four successive nights
  by Mauna Loa (left) and Haleakala (right).  The
  bottom row shows the cumulative coverage for each summit.  Color codes
  the number of visits: blue through red is 1--4 visits.  Gaps arise
  because of imperfect weather, duplicates because of morning clouds
  in the east compelled the schedule to widen Dec coverage.}
\label{fig:decbands}
\end{figure}

Figure~\ref{fig:decbands} shows the ATLAS sky coverage during a recent set of
four nights.

There were many motivations to site both of the first two ATLAS units
in Hawaii on Haleakala and Mauna Loa, one of which is weather
diversity.  Over the 122 days of Jun--Sep 2017, there were only 10
nights when both summits produced fewer than 450 successfully
differenced images, i.e. 92\% of nights were at least half workable
for at least one summit.  For each individual summit about 80\% of
nights are at least half productive, illustrating a degree of
decorrelation of the weather.  Similarly the fraction of nights
during these 122 days for which the median zeropoint was within 0.1
mag of clear was 69\% on Haleakala and 89\% on Mauna Loa, and 92\% of
nights were clear on at least one summit.

We normally observe in $c$ band during dark time on Haleakala and $o$
band during bright time, and we have always observed in $o$ band on Mauna
Loa but will soon start using $t$.  This provides color information for all the asteroids in the sky
as well as other transients and variables, without compromising our
sensitivity to find asteroids.
Currently Haleakala can achieve a 1.9 pixel FWHM in $o$ band and has a
median of about 2.1 pixels.  This degrades to 2.4 and 2.5 pixels in
$c$ band, presumably from chromatic aberration, but the darker sky
compensates and the zeropoint is about the same.  The best 5$\sigma$
limiting magnitude we achieve in 
a 30 second exposure 
is 19.8 (a zeropoint of 19.5 is common), and the
median over all lunations and sky conditions is 19.12.  On Mauna Loa
the sensitivity is just about 0.3 magnitudes worse because of the
degraded PSF, and we can achieve 19.4, we often achieve 19.2, and the
median is 18.83.

\begin{figure}[H]
\begin{center}$
\begin{array}{ccc}
\includegraphics[height=2.5in]{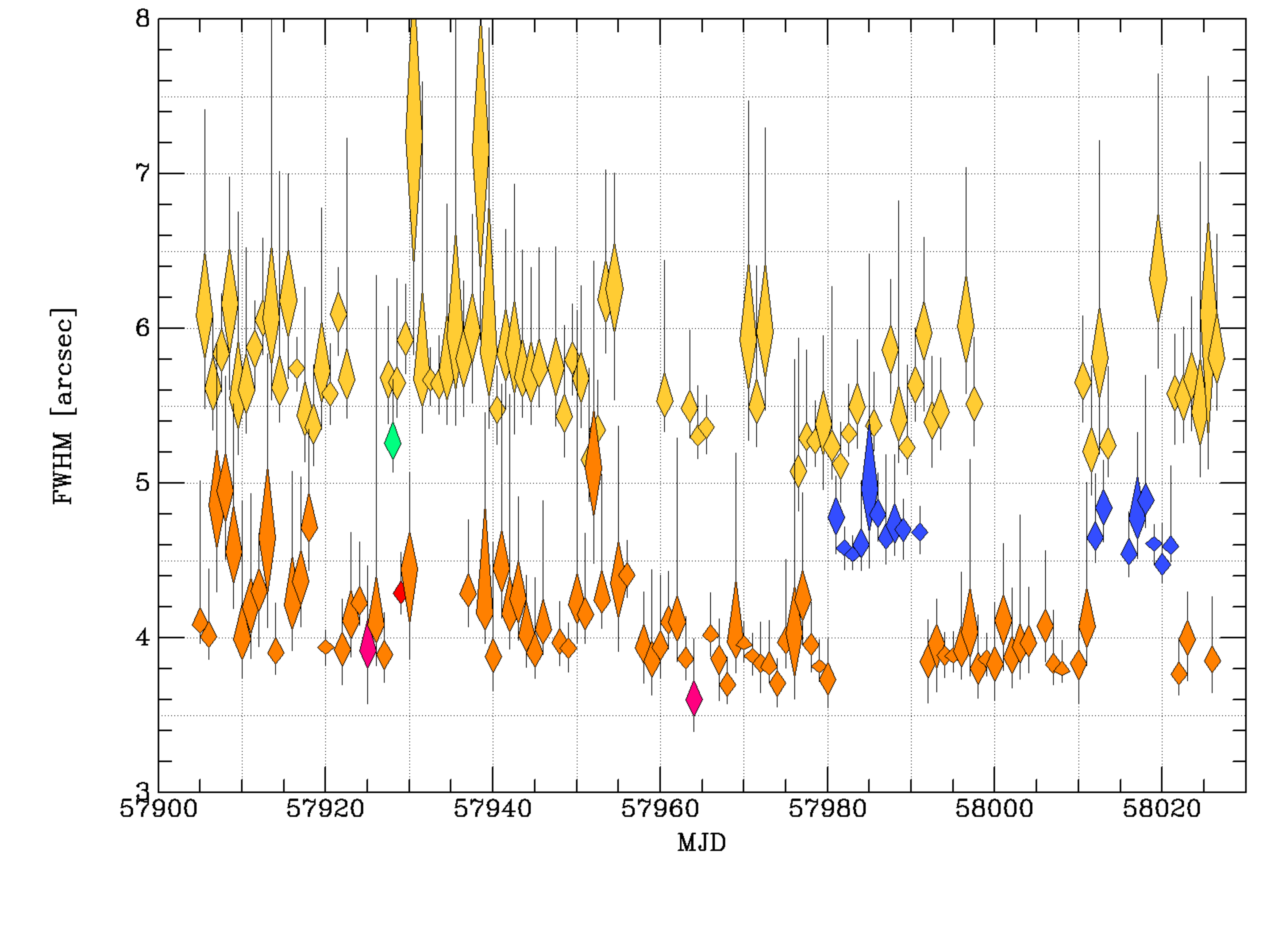}
\qquad
\includegraphics[height=2.5in]{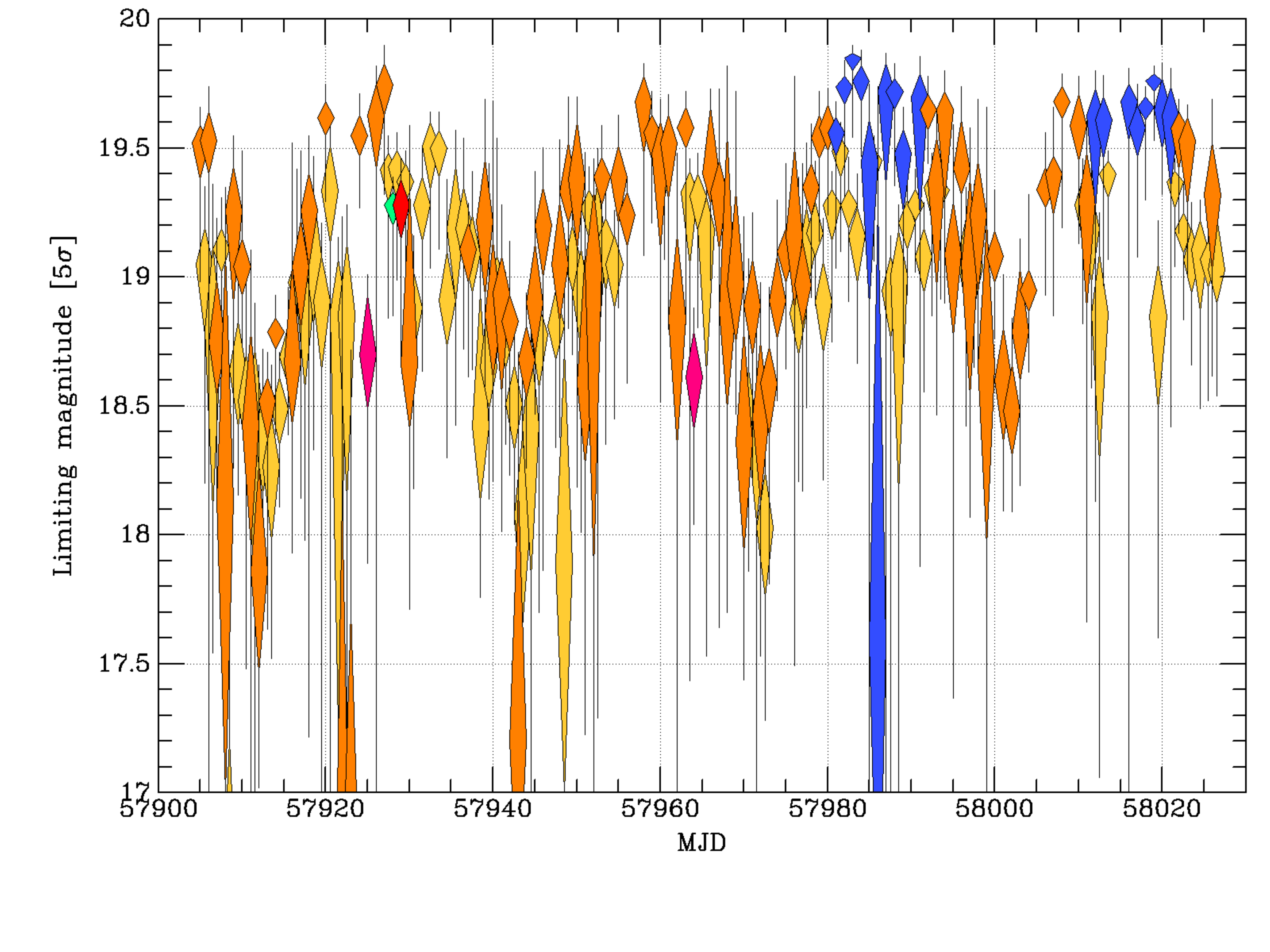}
\end{array}$
\end{center}
\caption{The left image shows the FWHM as a function of time and the
  right shows the limiting magnitude.  The diamond symbols show the
  median and quartiles for each night, with lines extending to the 5\%
  extremes of the distributions.  The filter and site determine the
  point color: yellow is MLO in $o$ band, orange is HKO in $o$ band,
  blue is $c$ band, green is $g$, red is $r$, and magenta $i$.  The
  $gri$ exposures from HKO are part of an ongoing effort to improve
  our photometric catalog south of $-30\deg$ Dec.
  The
  ``limiting magnitude'' is defined by where dophot reports a median
  uncertainty of 0.2 mag; the effect of sky brightness during four
  lunations is evident as well as cloudy nights.}
\label{fig:fwsig}
\end{figure}

The FWHM and limiting magnitude performance over a recent,
representative 4 month period (Jun-Sep 2017) are shown in
Figure~\ref{fig:fwsig}.  It is an important priority for us to
understand and improve the PSF in order to realize the full potential
of our system: we are making progress but work remains.
To disentangle seeing effects from defocus or mount shake we
have installed a 125~mm, f/20 refractor on MLO and we have started
collecting observations simultaneous with every science exposure.

\begin{figure}[H]
\begin{center}
\includegraphics[width=4in]{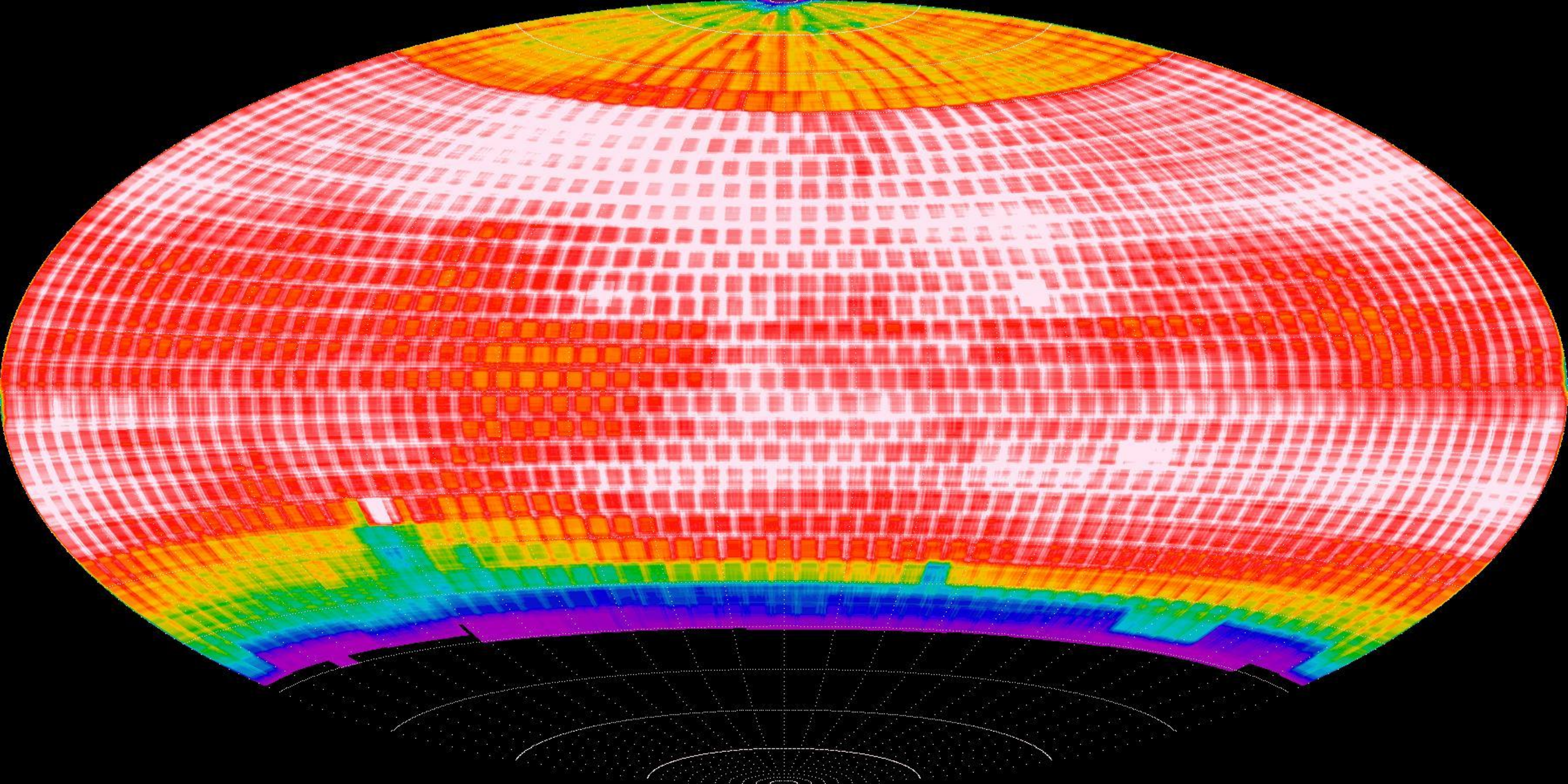}
\end{center}
\caption{Approximately 600,000 photometrically calibrated
  ATLAS exposures taken between Sep 2015 and Jan 2018 are shown as a
  function of equatorial coordinates (left to right is RA 360\deg--0\deg); the number of
  visits coded by color: cyan is 100, yellow is 300, red is 500, and
  white is more than 600.  There 240 million stars with at least 100
  light curve points from this coverage.}
\label{fig:skycover}
\end{figure}

Figure~\ref{fig:skycover} shows the number of times each spot on the
sky has been observed by ATLAS to date.  These include both
$c$ and $o$ filters, but only observations that are photometrically
calibratable.  By tying to Pan-STARRS1 reference stars 
\citep{magnier2017c}
most
observations north of Dec $-30\deg$ should have photometric errors at the
0.01 mag level or lower.  South of Dec $-30\deg$, APASS \citep{2012JAVSO..40..430H} photometry is good
to about 0.05 mag in systematics.  We are in the process of gathering
our own photometry to fill in $g$, $r$, and $i$ reference stars down
to Dec $-50\deg$.  Using Gaia astrometry, there should be negligible
systematic error in
the positions, and the median RMS astrometric error for stars
brighter than $m=17$ is about 70~milliarcsec; for fainter objects the error increases
inversely as SNR. 
The release of SkyMapper's First Data Release \cite{2018arXiv180107834W} 
provides another source of calibration for us to use below Dec $-30\deg$,
and we sill soon build a new reference catalog incorporating Gaia DR2, Pan-STARRS, APASS,
SkyMapper, and ATLAS photometry.

As of the end of January 2018 ATLAS has discovered 125 NEAs, 16 PHAs, and (somewhat
surprisingly) 9 comets, and we have submitted 5.5 million
observations of 128,000 distinct asteroids to the MPC.  
This capability is a direct result of efforts to
improve sensitivity and characterization for trailed asteroids on the
ATLAS detectors.  ATLAS participated in the October 2017 flyby exercise of asteroid 2012 TC$_4$ 
and was able to detect TC$_4$ during routine operations three days
before its close approach, despite the asteroid being very close to
the waning full moon.  ATLAS observations were submitted normally,
posted on the MPC confirmation page, and even flagged as ``very
close'' by JPL Scout hazard assessment service\footnote{\tt https://cneos.jpl.nasa.gov/scout/} before it was confirmed to be TC$_4$.

\begin{figure}[h]
\begin{center}
\includegraphics[height=3in]{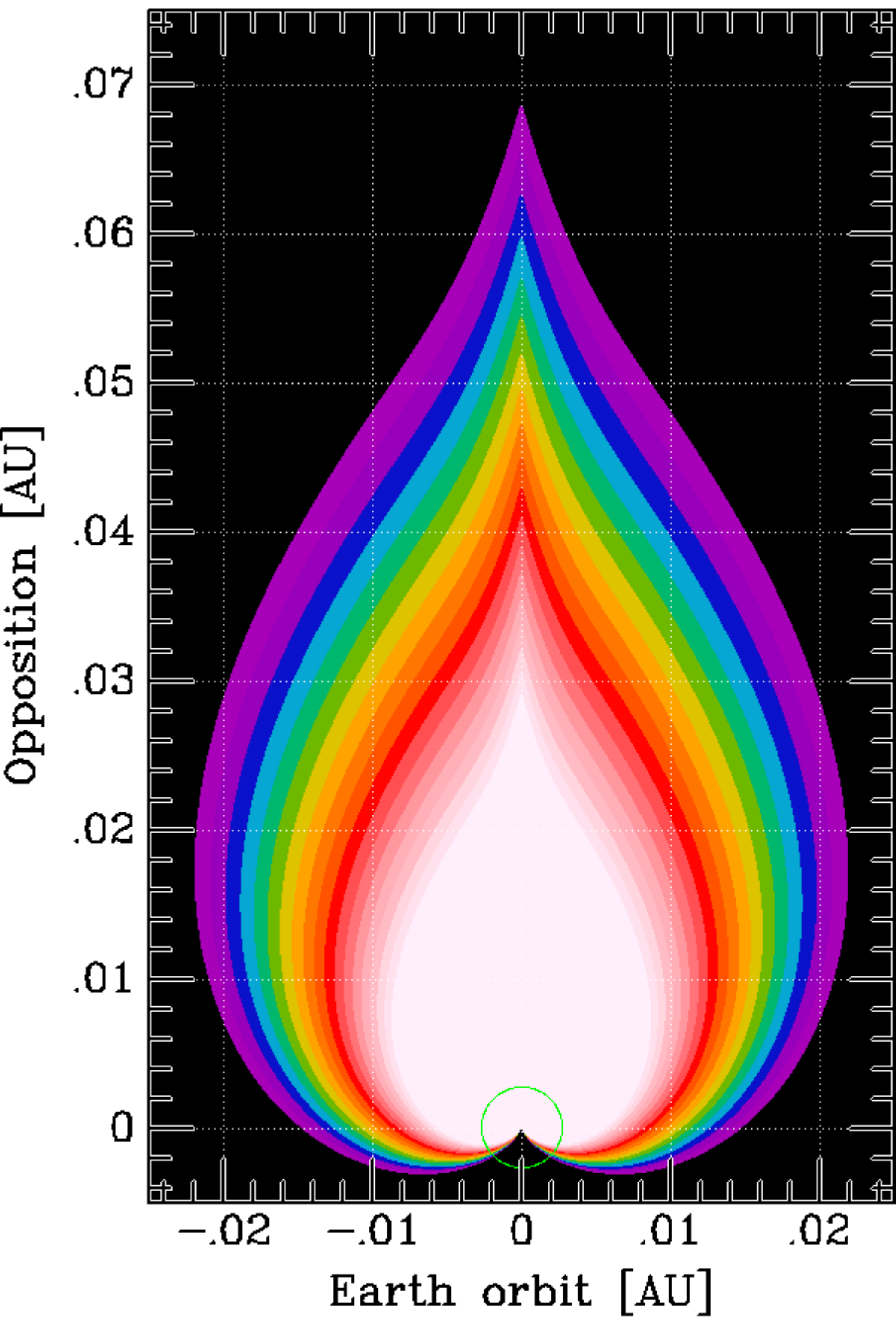}
\end{center}
\caption{The volume in which a 30~m diameter asteroid with typical
  phase function and albedo would be visible to ATLAS is shown with color coding
  the probabilty that the asteroid would be seen: black is 0.0, yellow is 0.5, and white is 1.0.
  The Moon's orbit is a green circle, and the Sun is at $(0,-1)$~AU.
  A typical asteroid will cross 0.01~AU in a day.  The 
  probabilities are derived from the actual performance of Haleakala
  during a recent, representative four month span; moon and clouds
  cause the variation in probability and volume.}
\label{fig:tulip}
\end{figure}

One of the more important features of ATLAS for the overall NASA NEOO
program is the ability to find very nearby objects, follow them, and
report them quickly.  Figure~\ref{fig:tulip} shows the
``candle flame'' volume accessible to ATLAS for an asteroid of diameter
30~m. The probability of entering
this volume is 80,000 times greater than the probability of striking
the Earth.  Depending on whether the near Earth population is normalized by
Earth impact rates estimated by
\cite{2002Natur.420..294B} or \cite{2013Natur.503..238B}, the expectation for the number
of asteroids in this volume at any given instant is either 2 or 9, and
the refresh rate is approximately 5 days.  Given full sensitivity to
streaked detections and allowing for moon and clouds, ATLAS should be
able to see about 10--50 NEAs of size 30~m per year, depending on
which of these normalizations is correct.

\begin{figure}[h]
\begin{center}
\includegraphics[width=4in]{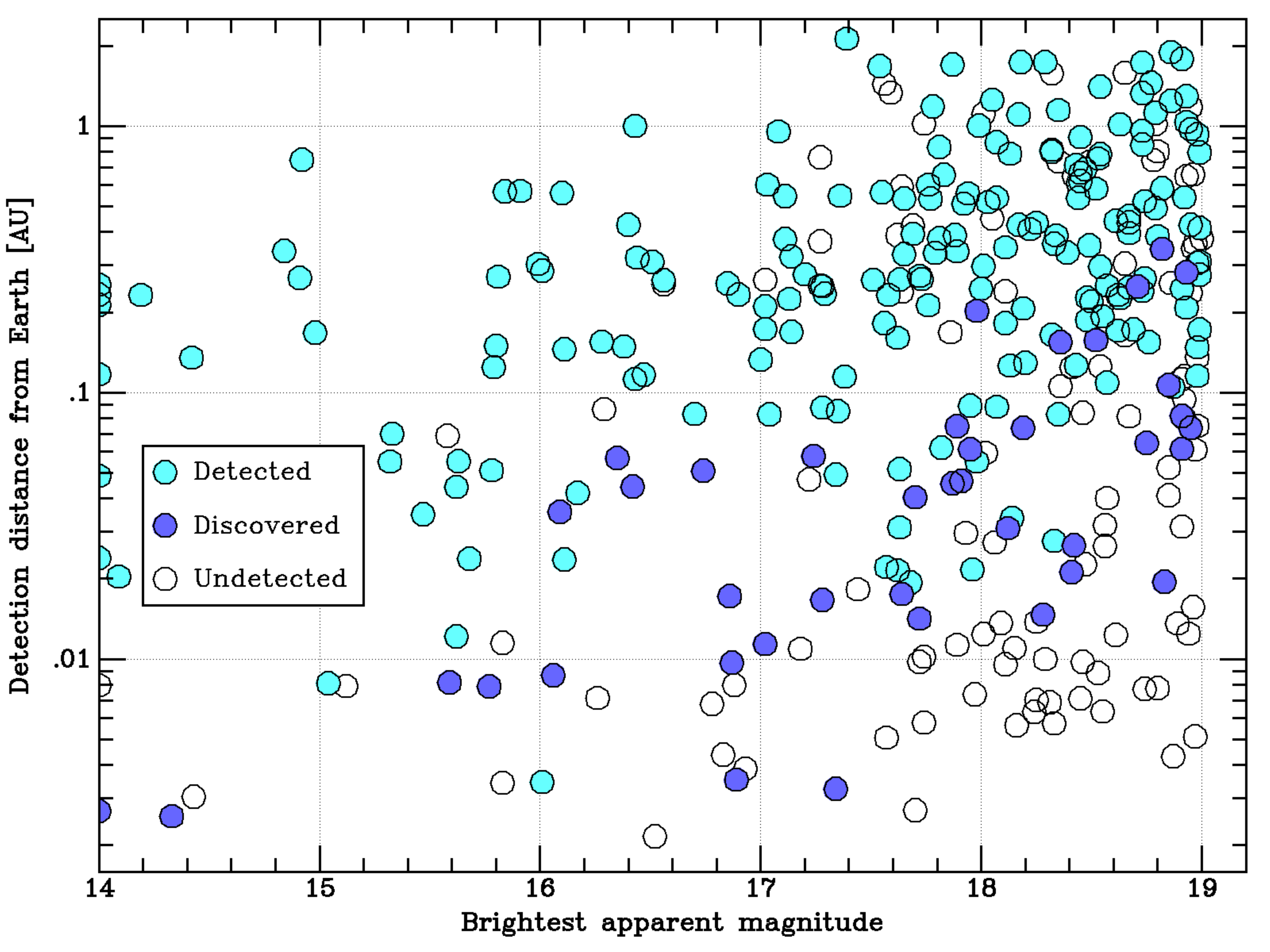}
\end{center}
\caption{ATLAS detected NEAs as a function of Earth distance over the 122-day interval
01 June 2017 through 30 September 2017.  The magnitude either refers to ATLAS $c$ or $o$ for
ATLAS detections or else $V$ band. The white points are NEAs that
were either previously known or discovered by another survey first during this period, but not
detected by ATLAS.  Cyan and blue points are NEAs that were detected by ATLAS, with blue
indicating that ATLAS was the discoverer.  ATLAS is very efficient at detecting
nearby asteroids that are brighter than $m<19$, and is particularly effective for discovering NEAs closer than 0.1~AU.}
\label{fig:atlasneos}
\end{figure}

NEA detection and discovery statistics shown in Figure~\ref{fig:atlasneos} illustrate ATLAS' successful optimization for its mission of detecting asteroids passing near the Earth. While other surveys with larger telescopes and greater sensitivity discover many more asteroids per year, ATLAS' very rapid sky coverage gives it a unique ability to detect almost all the asteroids that brighten past its sensitivity limit in any given interval of time. During the period covered by Figure~\ref{fig:atlasneos} (June-September 2017), ATLAS detected 75\% of all NEAs determined by after-the-fact ephemerides to have brightened past magnitude 19.0 at declination north of $-35$\deg\ and solar elongation greater than 90\deg. The 287 NEAs used to calculate this statistic include challenging objects that remained above the 19th magnitude detection threshold for less than one day; became observable only near full Moon; or were moving as fast as 50 deg/day at the moment of their closest approach. Of the 75\% of sufficiently bright NEAs that were detected by ATLAS, 19\% (40 NEAs) were ATLAS discoveries, and most of these would not even be known to exist or to have passed Earth apart from ATLAS. The 25\% of potentially detectable NEAs that ATLAS missed includes some that were discovered earlier and not recovered by any survey during their 2017 apparition. Thus, within a generous range of parameters that includes very difficult cases, NEAs passing Earth had only a 25\% chance of escaping ATLAS' net in late 2017. 

This 25\% statistic does not, of course, include an unknown number of NEAs that passed by Earth without being discovered by \textit{any} survey. The total number of small yet dangerous NEAs that inhabit the Solar System is still uncertain to within a factor of a few -- but ATLAS is poised to measure it much more accurately over the coming years.

In this context, we can consider ATLAS' ability to detect asteroids in the size range of the object that likely produced the Tunguska explosion in 1908, which devastated more than 2000 km$^2$ of Siberian forest. This object likely had an absolute magnitude of $H\sim25$, which corresponds to a diameter of 60m for a 5\% albedo. We like to use the term `Tunguska-level near miss' to describe cases when asteroids at least this large are discovered during Earth encounters that bring them closer than 0.01 AU. Neglecting gravitational focusing, there should be one actual Tunguska-like impact for every 55,000 Tunguska-level near misses. During the June-September 2017 period covered by Figure~\ref{fig:atlasneos}, there were seven known Tunguska-level near misses, which may be naively translated into an impact rate of one per 3000 years. Of the seven near-misses, ATLAS discovered three and Catalina and Pan-STARRS each discovered two. Thus, although the larger telescopes of the other surveys enable them to discover far more NEAs than ATLAS overall (including larger, globally hazardous asteroids passing Earth at larger distances), in the specific case of regionally dangerous asteroids passing very close to the Earth, ATLAS is competitive with far more expensive surveys. We note also that six out of the seven Tunguska-level near misses were detected by only one of the major surveys. This lack of overlap suggests that only a minority of the actual Tunguska-level near misses are detected by any survey -- a gap in our current planetary defence that can be most efficiently filled by building more ATLAS units. When the NEA population statistics are finally determined, the rate of Tunguska-like impacts will likely turn out to be considerably higher than the naive estimate quoted above.

\begin{figure}[h]
\begin{center}$
\begin{array}{ccc}
\includegraphics[height=2.5in]{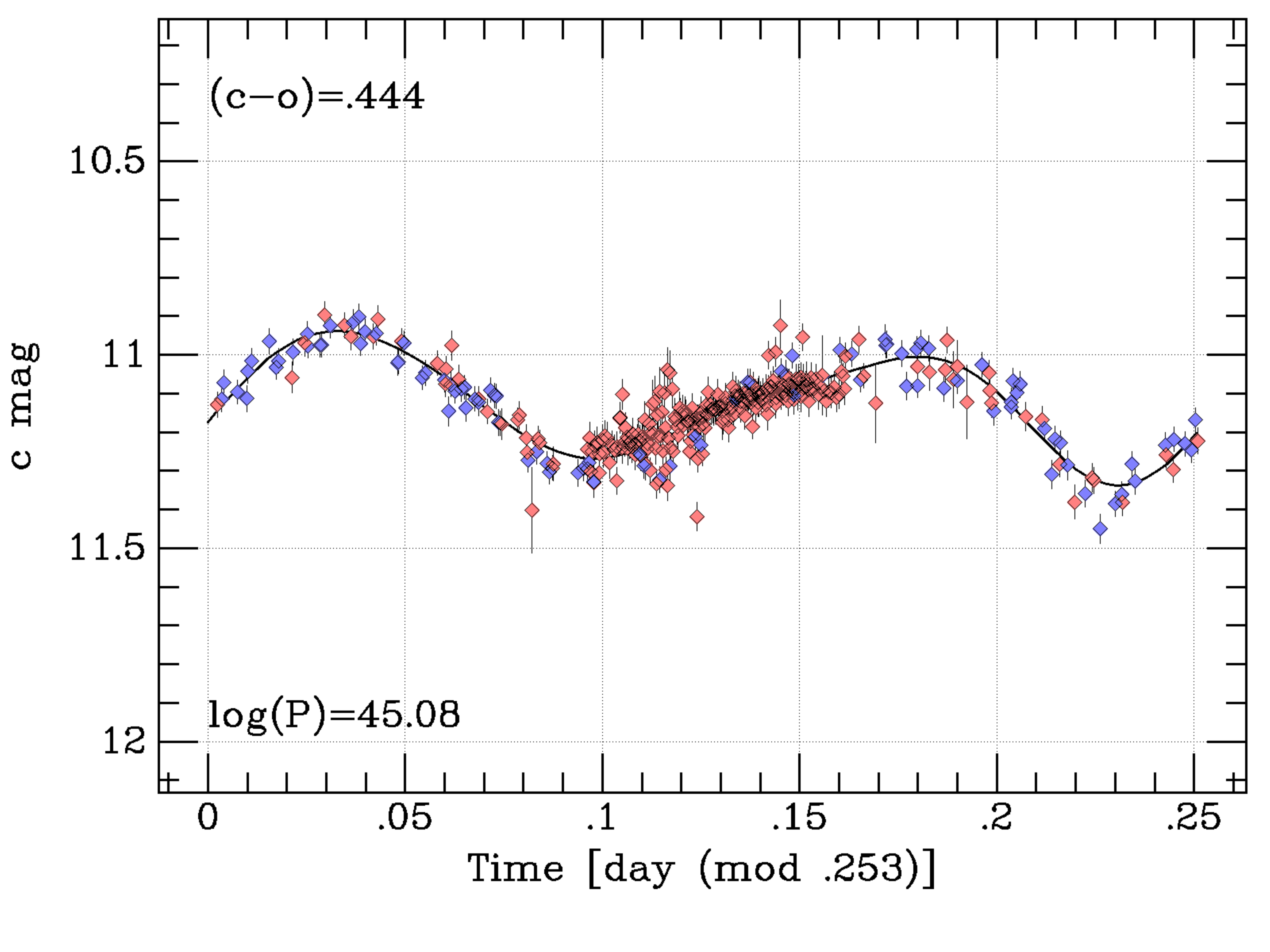}
\qquad
\includegraphics[height=2.5in]{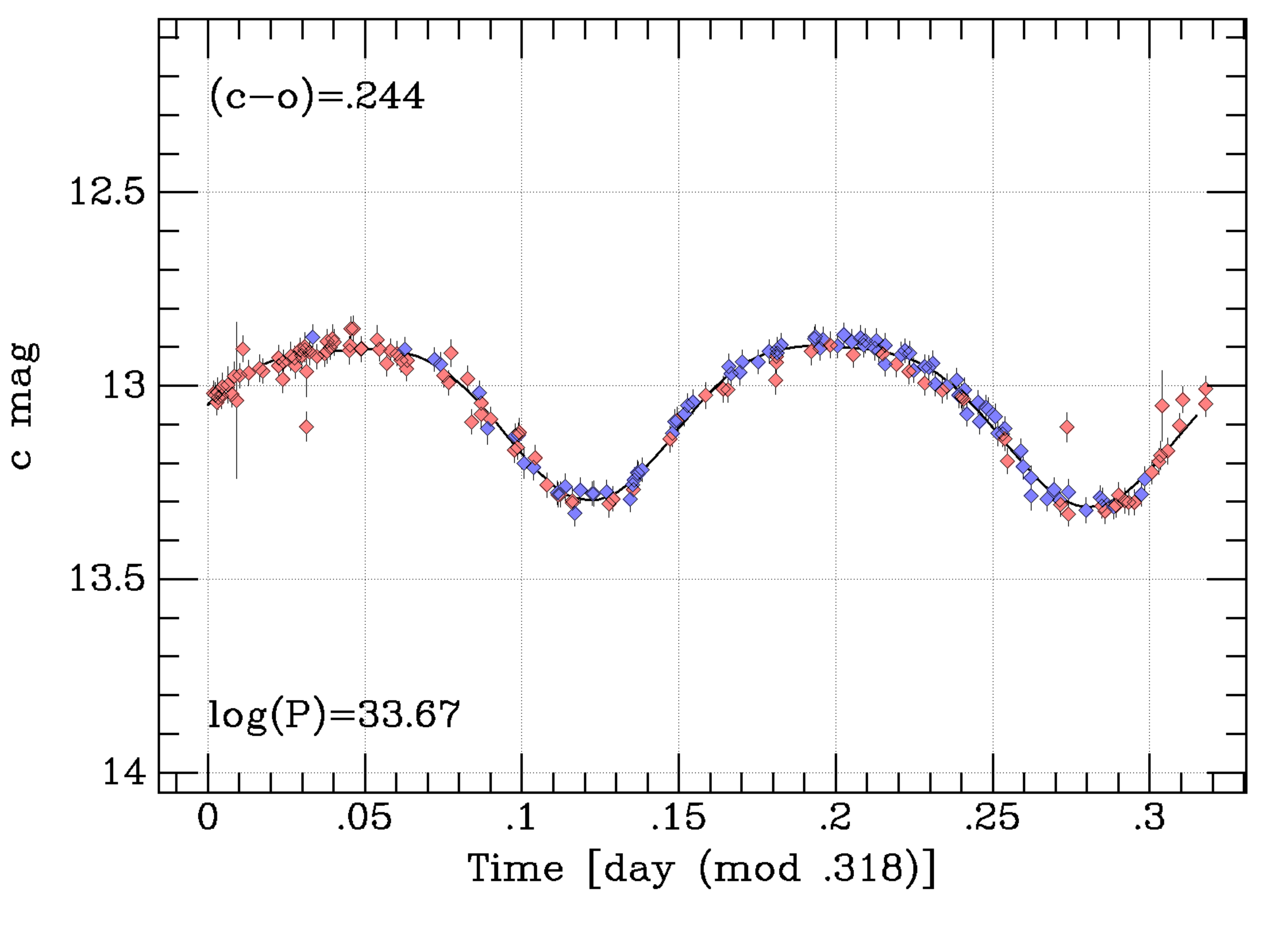}
\end{array}$
\end{center}
\caption{The left image shows the H magnitude phased light curve for
  the bright asteroid 700, and the right image a bright, AAVSO
  eclipsing binary star.  Point color indicates $c$ (blue) or $o$
  (orange) filters; the $o$ magnitudes are shifted by the listed $(c-o)$ color.
  ``log(P)'' is $-\log$ of the false alarm probability reported by Lomb-Scargle.
  The ATLAS sampling and coverage can recover an accurate period as well as color, even
  though the observations span many periods.  ATLAS has lightcurves
  for 240 million other stars and asteroids.}
\label{fig:kast}
\end{figure}

\begin{figure}[h]
\begin{center}$
\begin{array}{ccc}
\includegraphics[height=2.5in]{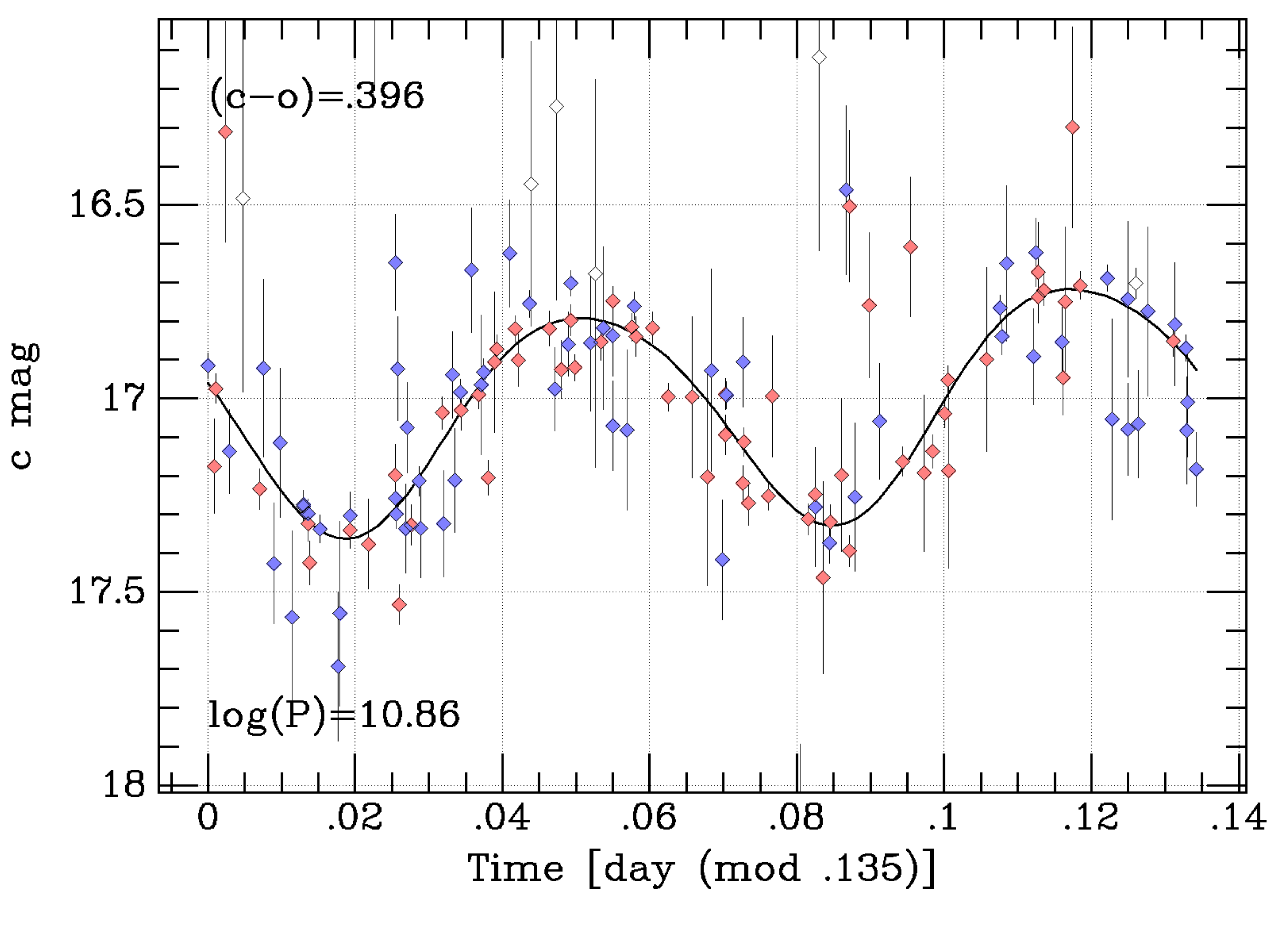}
\qquad
\includegraphics[height=2.5in]{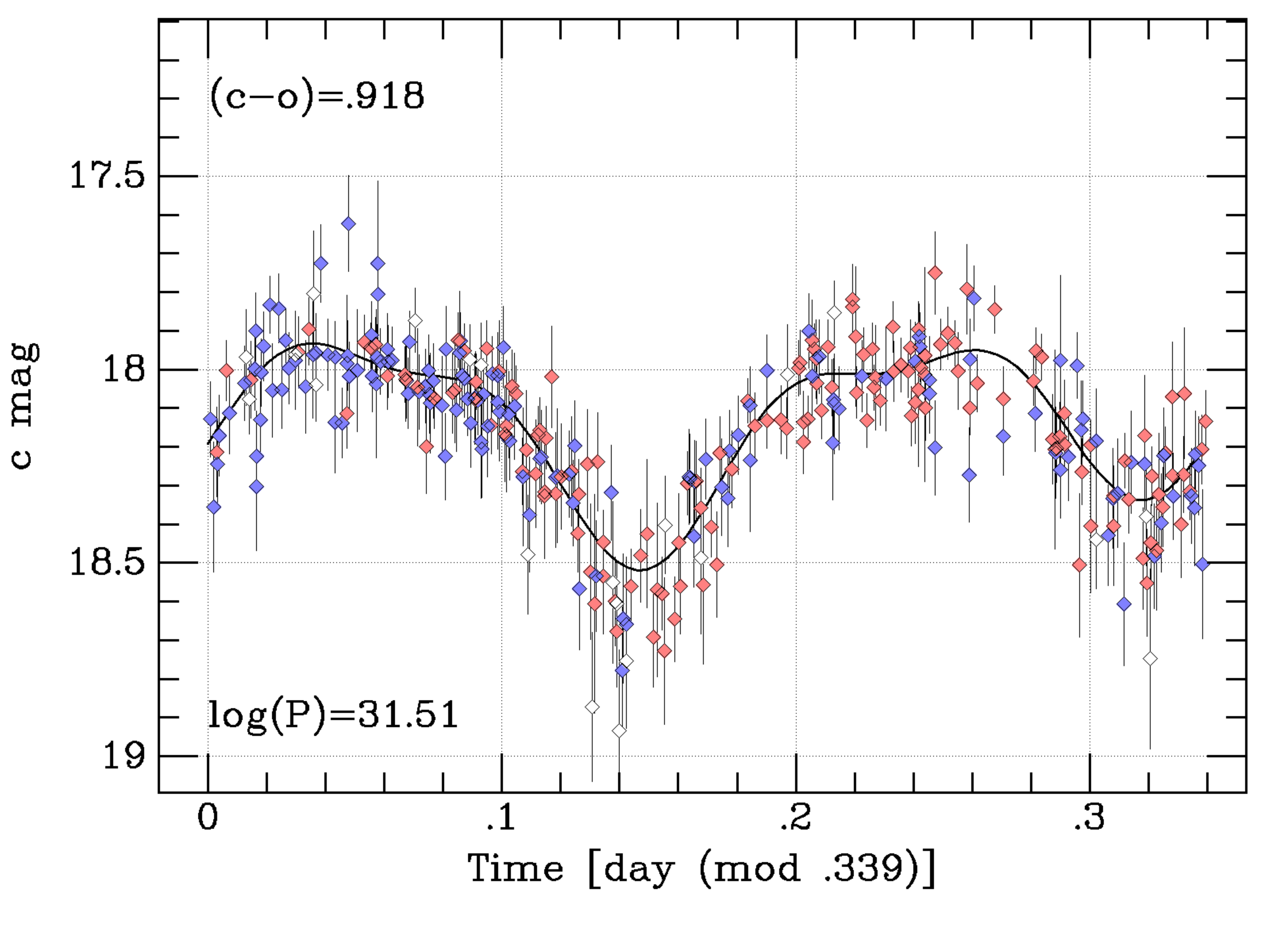}
\end{array}$
\end{center}
\caption{The left image shows the H magnitude phased light curve for
  the near Earth asteroid 1997\_XF11 (35396) that rotates every 3.26
  hours, and the right image an uncataloged eclipsing binary star
  with a period of 8.15 hours.  White points are deemed outliers and not included by our Lomb-Scargle program.}
\label{fig:dimkast}
\end{figure}

Figure~\ref{fig:kast} shows lightcurves for a typical bright asteroid
and a typical bright variable star.  
Figure~\ref{fig:dimkast} shows lightcurves for an asteroid that is
15 times smaller and a variable star that is 100 times fainter.
ATLAS has analyzed 2 year lightcurves
with at least 100 points for about 20,000 asteroids and 140 million
stars between $-30\deg<\delta<+60\deg$ and $m<18$, and detected 5 million 
stars which show variability  (Heinze et al. {\em in prep.}).
These lightcurves will be available from MAST at Space Telescope Science Institute.
Observation to date has
increased the number of stars to 240 million between $-45\deg<\delta<+90\deg$.

Our improved photometry enables light curve science with ATLAS data.
We have observations of $\sim$300,000 numbered asteroids, and as of 2017
(prior to replacement Schmidt correctors) we were able to assign
a period to 20,000 of them using standard period searches followed
by deep machine learning using the R programming language to classify
the candidate lightcurves.  This dataset alone is as large as the
current published asteroid lightcurve database and includes
thousands of previously unknown asteroid lightcurves.  Continued ATLAS
observations will only improve the number and quality of asteroid
lightcurves in the ATLAS dataset.

\begin{figure}[h]
\begin{center}$
\begin{array}{ccc}
\includegraphics[height=2.0in]{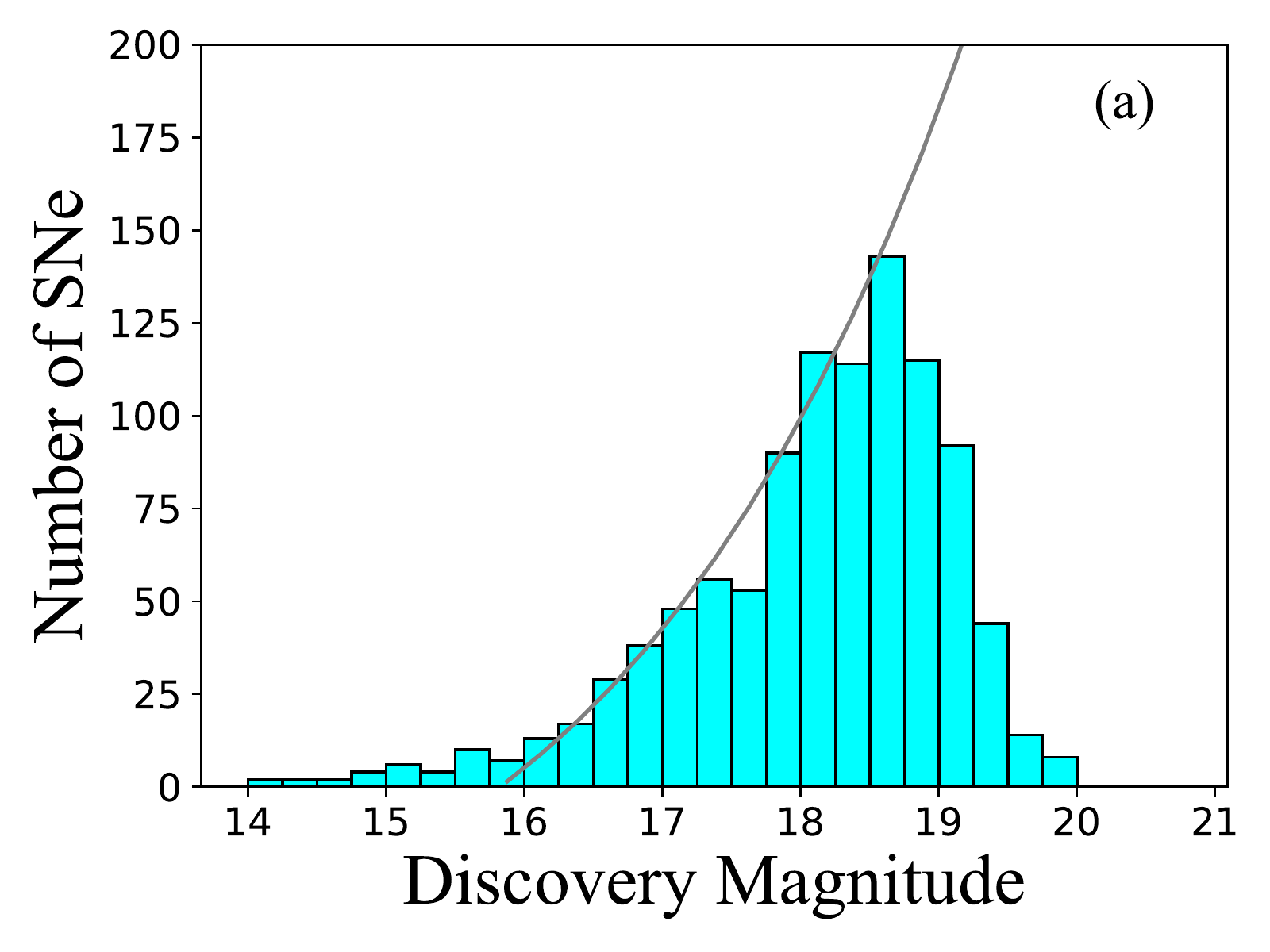}
\qquad
\includegraphics[height=2.0in]{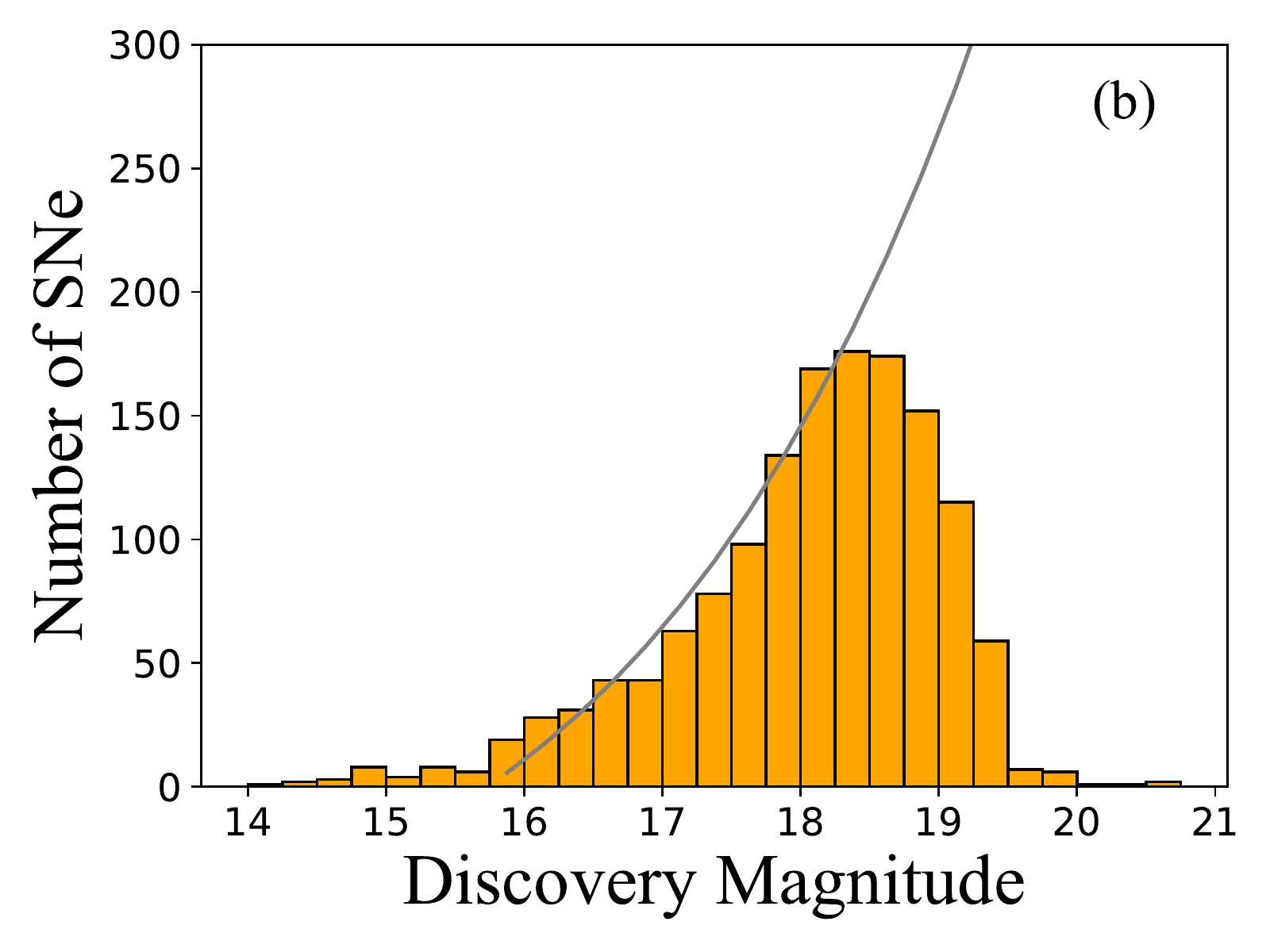}
\end{array}$
\end{center}
\begin{center}$
\begin{array}{ccc}
\includegraphics[height=2.0in]{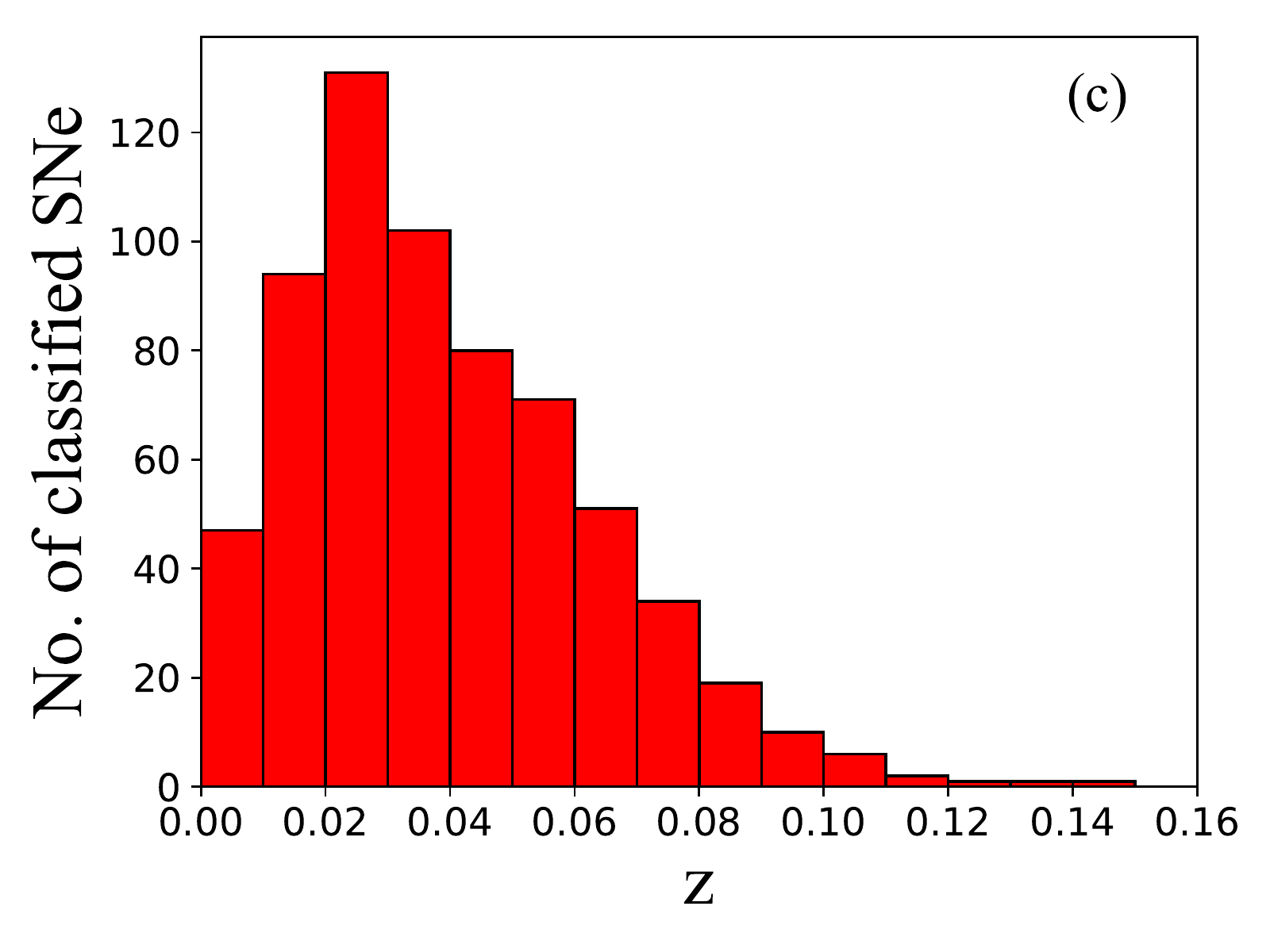}
\qquad
\includegraphics[height=2.0in]{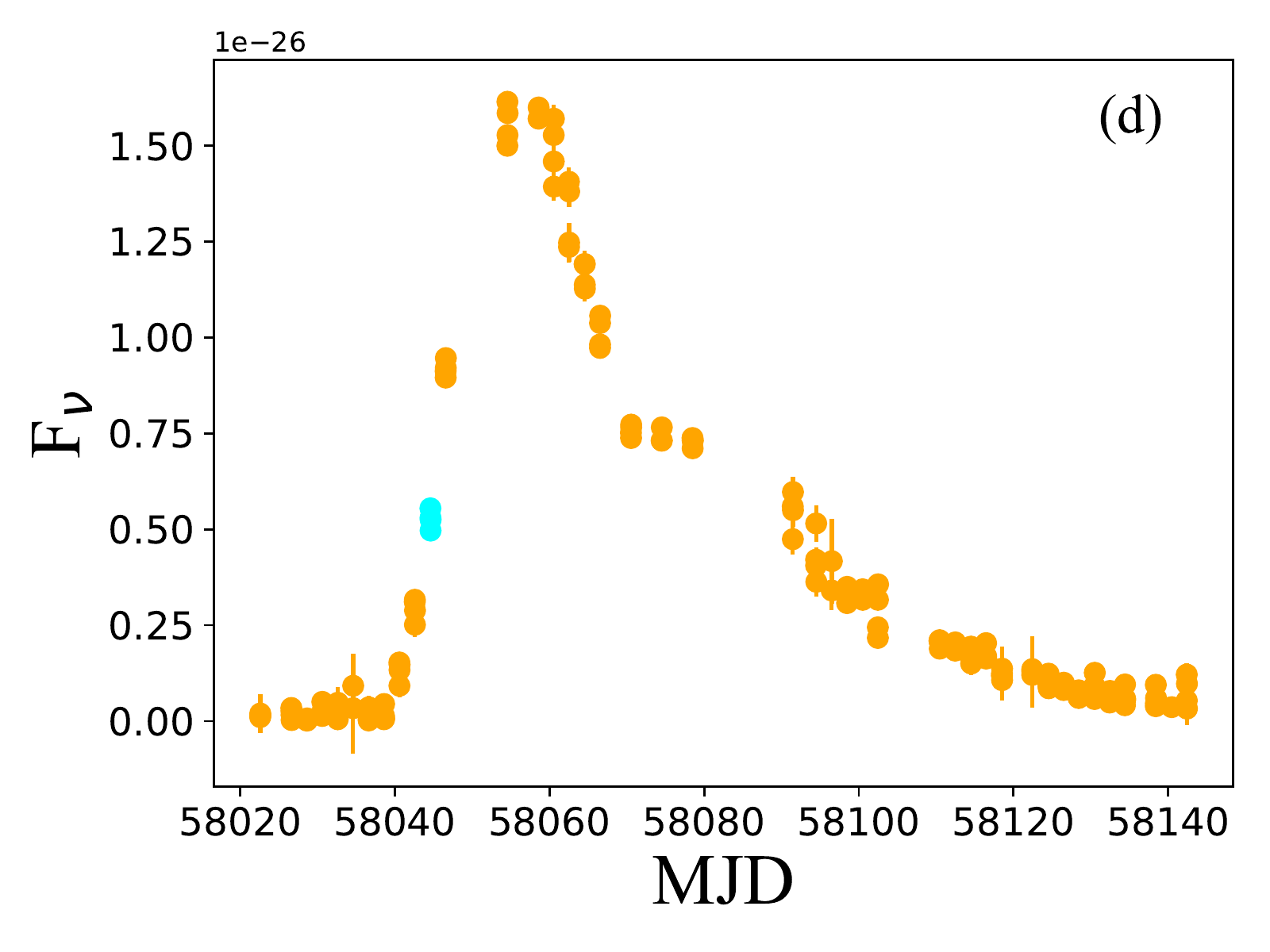}
\end{array}$
\end{center}
\caption{{\bf(a):} histogram of discovery magnitudes of the 1075 ATLAS detected SNe in the cyan filter. This includes our own discoveries, and those discovered by other surveys, but which are independently detected in our survey. 
{\bf(b):} histogram of discovery magnitudes of the 1441 ATLAS detected SNe in the orange filter. 
{\bf(c):} histogram of redshifts of the 654 classified SNe (only a small fraction of discoveries receive the followup required for secure classification). 
{\bf(d):} forced photometry (flux vs epoch) of SN2017hjw at $z=0.016$, obtained from the 
normal observations seeking asteroids.
}
\label{fig:snsummary}
\end{figure}


ATLAS has been a partner in LIGO - Virgo Consortium searching for optical
counterparts for gravitational wave events 
\citep{2016ApJ...826L..13A,2017ApJ...848L..12A,2017ApJ...850..149S}
The ATLAS survey is particularly
well-suited because our system covers such a large sky area that we 
do {\it not} require targeted scheduling
to cover areas of interest when LIGO events occur. We can provide 
the history of variablity or transient activity in LIGO-Virgo skymap 
to reduce the astrophysical false positives and reject associations. 
The one electromagnetic counterpart discovered so far, AT2017gfo 
(associated with GW170817), was remarkably bright when discovered at 0.47\,days 
after the GW trigger. From the early $gri$ photometry of \cite{2017Natur.551...64A,2017arXiv171005443D,2017arXiv171005452C}, we estimate its peak 
brightness within the first day 
was $c\sim17.4$ and  $o\sim17.3$, comfortably above the routine ATLAS survey
limits. We were unfortunate with the sky placing, since we had stopped observing 
that RA range just 16 days before. Up to that point ATLAS had 601 individual images
of NGC4993, providing a simple but important statement that no variable or transient object 
had ever been detected at that position and provided a temporal constraint for 
the 4-dimensional probability (3 space and one of time) 
of coincidence of the optical transient and the gravitational wave \citep[as discussed in][]{2017Natur.551...75S}. 
With our large sky area coverage, 2 day cadence and rapid data processing ATLAS will play 
an important role in the detection of future electromagnetic counterparts in LIGO-Virgo's O3 run
beginning 2018, and will also provide an independent search for 
kilonova without a GW trigger, within a volume of about 60\,Mpc. 

Figure\,\ref{fig:snsummary} shows the earliest ATLAS magnitudes of all
2476 
candidate extragalactic transients we have identified since
21 Dec 2015, the overwhelming majority
of which are supernovae. 
In this figure we do not distinguish
between those supernovae that were discovered (i.e. first report
to the TNS) by other surveys and those that were exclusively 
discovered first by ATLAS. The reason for doing this is to highlight the ATLAS detection performance, rather than speed and competitiveness of reporting. At $m=19\pm0.2$ we are roughly 50\% complete assuming that supernovae are isotropically distributed in the local volume. We also show an example of the lightcurve of an ATLAS supernova (ATLAS17mgh, recorded as SN2017hjw on the TNS, also seen by Gaia as Gaia17crm) with 
forced photometry (flux vs epoch). This is a type Ia in UGC03245
at  $z=0.016$ and  a spectrum was  
taken about 1 week before maximum light by the 
Asiago Transient Classification Program 
\citep{2014AN....335..841T,2017ATel10863....1T}, enabled by the immediate, public discovery announcement by ATLAS. The secondary peak in the lightcurve is nicely visible in the $o$-band filter. We also show the 
redshift distribution of the 654 spectroscopically classified 
supernovae in Figure\,\ref{fig:snsummary}. 
A full description of the ATLAS transient pipeline and early science results will be discussed in Smith et al. (in prep).



\section{Conclusions}

ATLAS represents another step in the relentless march toward increased time-domain coverage of the sky.  With ATLAS, the complete northern sky is now observed every two days to fainter than $m=19$.  The system routinely and automatically executes its mission of surveying for dangerous NEAs. The low cost and reproducibility of an ATLAS unit means that the system capability is relatively easy to extend.   

ATLAS stands on the shoulders of many other broad advances---Gaia for exquisite astrometry, Pan-STARRS for photometric calibration and its MOPS pipeline, {\tt astrometry.net} for blind astrometric reduction, to name several---all facilitated by inexpensive, high-performance computers.  Although we are not outfitted to serve ATLAS data products to the community, ATLAS data are available to any institution able to receive our data, with no proprietary period.

Prospects for extending the ATLAS system to cover the remaining southern sky are excellent -- there are plans awaiting funding to construct two ATLAS units in the southern hemisphere longitudinally opposite Hawai`i.  These additional units will allow the entire system to re-observe the entire sky every 24 hours (and every 12 hours for some of the sky).  The weather diversity and continuous sky coverage of such a configuration will improve the ATLAS dataset in areas where it already excels:

\begin{itemize}
  \item{} Tightening the net for NEA discovery.  While the current collection of all-sky NEA surveys (Pan-STARRS 1, Catalina Sky Survey and ATLAS) continue to increase their discovery rates, many detectable NEAs still go undiscovered each lunation.  A full-sky, nightly ATLAS system will reduce the number of undetected NEAs that sneak by the Earth, and the well-calibrated and characterized ATLAS system will help quantify global NEA survey effectiveness.
  \item{} Denser coverage for transients.  Detectable supernovae will be observed within 12-24 hours of explosion, increasing the likelihood of seeing shock breakout and obtaining followup spectra during the very interesting early stages of the explosion.
  \item{} New variable stars in challenging classes. These include stars with very small amplitudes; variables currently difficult to measure due to frequency aliasing such as RR Lyrae stars and contact binaries with periods near 0.5 or 1.0 sidereal days; and extremely long-period Mira stars and other pulsating supergiants.
  \item{} Immediate followup of any LIGO/Virgo transients.  The flexible asteroid survey can rearrange its target list such that an ATLAS telescope can be pointed at a candidate LIGO/Virgo event within 60 seconds.
\end{itemize}

\section{Acknowledgements}

We acknowledge useful discussions with Gaspar Bakos, Klaus Hodapp, Robert Jedicke, Eileen Ryan, Tim Spahr and Richard Wainscoat.  Support for this work was provided by NASA grant
NN12AR55G under the guidance of Lindley Johnson and Kelly Fast.  We acknowledge support for transient science exploitation from the EU FP7/2007-2013 ERC Grant agreement n$^{\rm o}$ [291222], STFC Grants ST/P000312/1, ST/N002520/1 and support from the QUB Kelvin HPC cluster, 
and the QUB International Engagement Fund. We thank Mike Bessell for discussions  on the 
design of our filter set and sharing the Skymapper experience.

\bibliographystyle{yahapj}
\bibliography{lib}


\end{document}